\documentclass[reprint,prd,twocolumn,superscriptaddress,nofootinbib]{revtex4-2}   % somehow this puts it on one page 
\usepackage{graphicx}
\usepackage{multirow}
\usepackage{gensymb}
\usepackage{amsmath}
\usepackage{bm}
\usepackage{natbib}
\usepackage{hyperref}
\usepackage{color}  %SdM: allows fancy color names
\newcommand\aj{{AJ}}%        % Astronomical Journal 
\newcommand\araa{{ARA\&A}}%  % Annual Review of Astron and Astrophys 
\renewcommand\apj{{ApJ}}%    % Astrophysical Journal 
\newcommand\apjl{{ApJL}}%     % Astrophysical Journal, Letters 
\newcommand\apjs{{ApJS}}%    % Astrophysical Journal, Supplement 
\newcommand\aap{{A\&A}}%     % Astronomy and Astrophysics 
\newcommand\mnras{{MNRAS}}%   % Monthly Notices of the RAS 
\renewcommand\prd{{Phys.~Rev.~D}}% % Physical Review D 
\newcommand\pasp{{PASP}}%     % Publications of the ASP 
\renewcommand\nat{{Nature}}%  % Nature 
\newcommand\rmxaa{{Rev. Mexicana Astron. Astrofis.}}%  % Revista Mexicana de Astronomia y Astrofisica

\def\mone{{\rm M}_1}

\def\zmax{z_{\rm max}}

\def\zevnt{z_{\rm event}}

\def\porb{P_{\rm orb}}

\def\tend{T_{\rm end}}
\def\nend{N_{\rm end}}
\def\nbhend{N_{\rm BH, end}}
\def\vcirc{V_{\rm circ}}
\def\taucl{\tau_{\rm cl}}
\def\mcomp{M_{\rm comp}}
\def\kc{\kappa_{\rm c}}
\def\kf{\kappa_{\rm f}}
\def\trh{\tau_{\rm rh}}
\def\etamrg{\eta_{\rm mrg}}
\def\sfr{\Psi(r_g)}
\def\sgas{\Sigma_{\rm gas}}

\newcommand{\Ms}{\ensuremath{{\rm M}_{\odot}}}

\newcommand{\eg}{{\it e.g.}}
\newcommand{\cf}{{\it c.f.~}}
\newcommand{\ie}{{\it i.e.}}

\newcommand{\beq}{\begin{equation}}
\newcommand{\eeq}{\end{equation}}

\newcommand{\kmps}{\ensuremath{{\rm~km~s}^{-1}}}

\newcommand{\thub}{\ensuremath{t_{\rm Hubble}}}

\newcommand{\mcl}{\ensuremath{M_{\rm cl}}}
\newcommand{\rh}{\ensuremath{r_{\rm h}}}

\newcommand{\rhbh}{\ensuremath{r_{\rm h,BH}}}
\newcommand{\rhns}{\ensuremath{r_{\rm h,NS}}}
\newcommand{\rhlum}{\ensuremath{r_{\rm h,lum}}}

\newcommand{\tmrg}{\ensuremath{t_{\rm mrg}}}

\newcommand{\tej}{\ensuremath{t_{\rm ej}}}
\newcommand{\vej}{\ensuremath{v_{\rm ej}}}

\newcommand{\nbseven}{{\tt NBODY7}}

\newcommand{\bse}{{\tt BSE}}

\newcommand{\archain}{{\tt ARCHAIN}}

\newcommand{\fbin}{\ensuremath{f_{\rm bin}}}

\newcommand{\fobin}{\ensuremath{f_{\rm Obin}}}

\newcommand{\mbh}{\ensuremath{M_{\rm BH}}}

\newcommand{\fmrg}{\ensuremath{f_{\rm mrg}}}
\newcommand{\ftz}{\ensuremath{f_{\rm TZ}}}

\begin{document}

\title[Open clusters in a tidal field]
{Stellar-mass black holes in young massive and open stellar clusters – VI. Role of
external galactic field}

\author{Sambaran Banerjee}
\email{sambaran@astro.uni-bonn.de (he/him/his)}
\affiliation{Helmholtz-Instituts f\"ur Strahlen- und Kernphysik (HISKP),
Nussallee 14-16, D-53115 Bonn, Germany}
\affiliation{Argelander-Institut f\"ur Astronomie (AIfA), Auf dem H\"ugel 71,
D-53121, Bonn, Germany}

\date{\today}

\begin{abstract}
	Young massive clusters (YMC) and open clusters (OC) are widely considered as potential
	environments for assembling merging binary stellar-remnant black holes (BBH) via dynamical interactions. However,
	such moderate mass systems (typically, $~10^3\Ms-10^4\Ms$) are susceptible to being disrupted
	by the external tidal field of their host galaxies, potentially limiting their effectiveness
	as gravitational-wave (GW) sources.
	In this study, I investigate the formation of BBH mergers in tidally dissolving
	star clusters. This is achieved with a newly computed grid consisting of 95 evolutionary model star clusters,  
	where the clusters are subjected to a varying extent of tidal stripping. The cluster
	evolutions are computed with the direct N-body integrator $\nbseven$ that includes, among others,
	treatments for post-Newtonian (PN) effects in compact-binary members,
	mass loss due to stellar evolution, formation of stellar remnants, and tidal stripping.
	It is found that even strong tidal stripping does not quench the formation of a black hole (BH) core
	inside a cluster or the formation of dynamical BBH mergers in the system. The overall properties of BBH
	mergers, \eg, the form of the distribution of merger delay time, primary mass, and mass ratio, and the redshift
	evolution of merger rate are not significantly altered by the extent of tidal stripping of the
	parent cluster population. Furthermore, even strongly tidally stripped clusters are capable of
	dynamically forming Gaia-BH-like detached BH--main-sequence-star binaries that escape into the galactic field. 
	Limitations of the present study and potential future improvements are discussed. 
\end{abstract}

\maketitle

\section{Introduction}\label{intro}

With mounting gravitational-wave (hereafter GW) events having been and being
detected by ground-based GW interferometers \citep{Abbott_GWTC3}, speculations regarding formation
mechanisms of these events have intensified (see Refs.~\cite{Mandel_2021,Spera_2022} for comprehensive reviews).
The majority of the to-date-observed GW events
are classified as binary stellar-remnant black hole (hereafter BBH) mergers. A diverse range of
scenarios have been proposed as potential formation channels for BBH mergers. Such scenarios include
dynamical interactions among stellar-remnant black holes (hereafter BH) in dense star clusters
\citep{Kulkarni_1993,Lee_1995,Benacquista_2013},
isolated evolution of massive-stellar binaries (\eg, \cite{Belczynski_2002}),
interacting BHs in AGN gas disks (\eg, \cite{McKernan_2018,Vaccaro_2024}),
isolated evolution of hierarchical massive-stellar systems (\eg, a massive field triple-star; \cite{VignaGomez_2021}),
close flyby interactions on BBHs in the galactic field (\eg, \cite{Michaely_2019}), and cluster and galactic
tides on wide BBHs (\eg, \cite{Hamilton_2019,Stegmann_2024}). Despite
being among the most extensively studied scenarios, there are aspects that warrant
further exploration in the physics of dynamical interactions of BHs inside dense stellar
systems.

Stellar-remnant black holes that receive natal kicks lower than the parent cluster's escape speed
would retain within the cluster and segregate close to the cluster's center to form
a dense core of BHs. Typically, only a $\lesssim10$\% BH retention fraction is sufficient
for the mass stratification instability to set in \citep{Spitzer_1987}. Therefore,
with a modest extent of BH retention, the BH core formation
is essentially a core collapse phenomenon, which would cause central energy generation and
expansion of the cluster, a process that is often referred to as
`balanced evolution' \citep{Henon_1975,Breen_2013,Antonini_2020}.
The phenomenon of expansion of a star cluster driven by energy generation inside
its BH core is also referred to as `black hole heating' (hereafter BH heating).

Energy is generated in the dense BH core through super-elastic close dynamical encounters
involving BBHs formed via the three-body mechanism \citep{Heggie_2003,Morscher_2015}.
Typically, such encounters are binary-single and binary-binary encounters among single BHs and BBHs
\cite{Antonini_2016b,Banerjee_2018,MarinPina_2024}.
The kinetic energy released in these scatterings is subsequently deposited onto the stellar background
of the parent cluster through dynamical friction, causing the cluster to expand \citep{Banerjee_2010,Wang_2016,Kremer_2020}.
The close encounters among the BHs and BBHs occasionally result in general-relativistic (hereafter GR) mergers
of the BBHs, either inside the cluster or after being dynamically ejected from the cluster
(often referred to as `in-cluster mergers' and `ejected mergers', respectively;
\cite{PortegiesZwart_2000,Banerjee_2010,Sippel_2013,Banerjee_2017,Park_2017,Rodriguez_2018,DiCarlo_2019,ArcaSedda_2024c}).
In contrast to BHs, retained neutron stars (hereafter NS),
owing to their significantly lower mass ($\gtrsim10\Ms$ vs $\approx1.3\Ms$;
\cite{Banerjee_2020}), are generally inefficient in heating the cluster
or producing GR mergers dynamically \citep{Ye_2020,Samsing_2021}.

A relatively unexplored aspect of BH dynamics in star clusters is the influence
of the external galactic gravitational field on a cluster's BH-heating-driven evolution
and GR merger production. In particular, while BH dynamics in young massive clusters (hereafter YMC)
evolving into open clusters (hereafter OC) is considered as a prospective formation channel for
the observed GR-merger events \citep{Banerjee_2017,Banerjee_2020c,Rastello_2019,Rastello_2021},
it is unclear how the galactic tidal forces would impact the channel's effectiveness. 
Typically having masses of $\lesssim10^4\Ms$ and parsec-scale size \citep{PortegiesZwart_2010,Banerjee_2018b,Krumholz_2019},
YMC/OCs are generally more susceptible to their host galaxy's tidal field compared to globular clusters (hereafter GC),
particularly when the former are formed in the galaxy's inner regions. 
Furthermore, unlike present-day GCs, present-day YMC/OCs potentially experience a
sustained tidal force from the host galaxy's disk as they would typically form inside and co-move 
with the galactic gas disk.

Efficient tidal stripping of a star cluster, containing a population of BHs, affects it in counteracting ways.
On one hand, the depleting stellar population decays the cluster's
potential-well and the background stellar density, which, otherwise, would have helped to keep the
BH-core compact and decelerate its dynamical decay \citep{Breen_2013,Heggie_2014}. On the other hand,
the preferential removal of the low-mass luminous members (low-mass stars and white dwarfs),
due to the stripping of the cluster's outer regions, makes the cluster BH-(and NS-)rich \citep{Banerjee_2011}
and accelerates its core collapse \citep{Spitzer_1987,Spurzem_1996,Banerjee_2010}. The former 
effect tends to inhibit BH heating and GR merger production while the latter effects aid
these processes. The goal of this work is to study these counter-playing aspects
simultaneously and in a consistent way.

To that end, a homogeneous grid of direct, star-by-star, relativistic N-body evolutionary
models of star clusters, covering a wide range of tidal radius filling, is computed and explored in this study
for the first time. Theoretically, the tidal radius or Jacobi radius, $r_t$, of a star cluster is
the effective radius of the critical Hill surface (or tidal lobe or Roche lobe) around the cluster in the cluster's host environment's
gravitational field. For the simplest case of a cluster of mass $\mcl$ orbiting a point-mass host galaxy of mass $M_g$
at a galactocentric distance $r_g$, the tidal radius is given by \citep{vonHoerner_1957}
\begin{equation}
	r_t = r_g\left(\frac{\mcl}{2M_g}\right)^{1/3}.
\label{eq:rt_pm}
\end{equation}
In the more general case of a cluster within a static mass distribution, $r_t$ is given by Eqn.~10 of
Ref.~\cite{Bertin_2008}. It then follows that as long as a cluster's orbit symmetrically encloses a constant mass
(\ie, the cluster always `sees' exactly the same mass distribution from its orbit),
\begin{equation}
	r_t \propto \mcl^{1/3}
\label{eq:rt_gen}
\end{equation}
to the lowest order.

Aspects of tidal dissolution of BH-population-containing star clusters have been investigated
by various authors. There has been a focus on formation of  
stellar-remnant-rich cluster remnants or `dark star clusters' and on
rapid tidal dissolution of clusters aided by BH heating
\citep{Banerjee_2011,Giersz_2019,Wang_2020,Wu_2024,Rostami_2024,Ghasemi_2024}.
Formation and properties of tidal tails are also of current interest
\citep{Weatherford_2024,Kroupa_2024,Roberts_2024}, so as observational
implications of a population of star clusters in a galactic field
\citep{Webb_2012,Webb_2021,Ishchenko_2025}.

This study, for the first time, focusses on BBH merger formation in a homogeneous evolutionary grid of
model star clusters that are significantly affected by both BH heating and
an external galactic field. Sec.~\ref{method} describes the
evolutionary model star clusters and the model grid.
Sec.~\ref{evolve} describes the evolutionary properties of these computed model clusters.
Sec.~\ref{gwmrg} discusses the GR mergers from the clusters.
Sec.~\ref{bhms} describes the formation of BH-star binaries from the models.
Sec.~\ref{summary} summarizes and further discusses the results. 

\begin{figure*}
\centering
\includegraphics[width = 8.9 cm, angle=0.0]{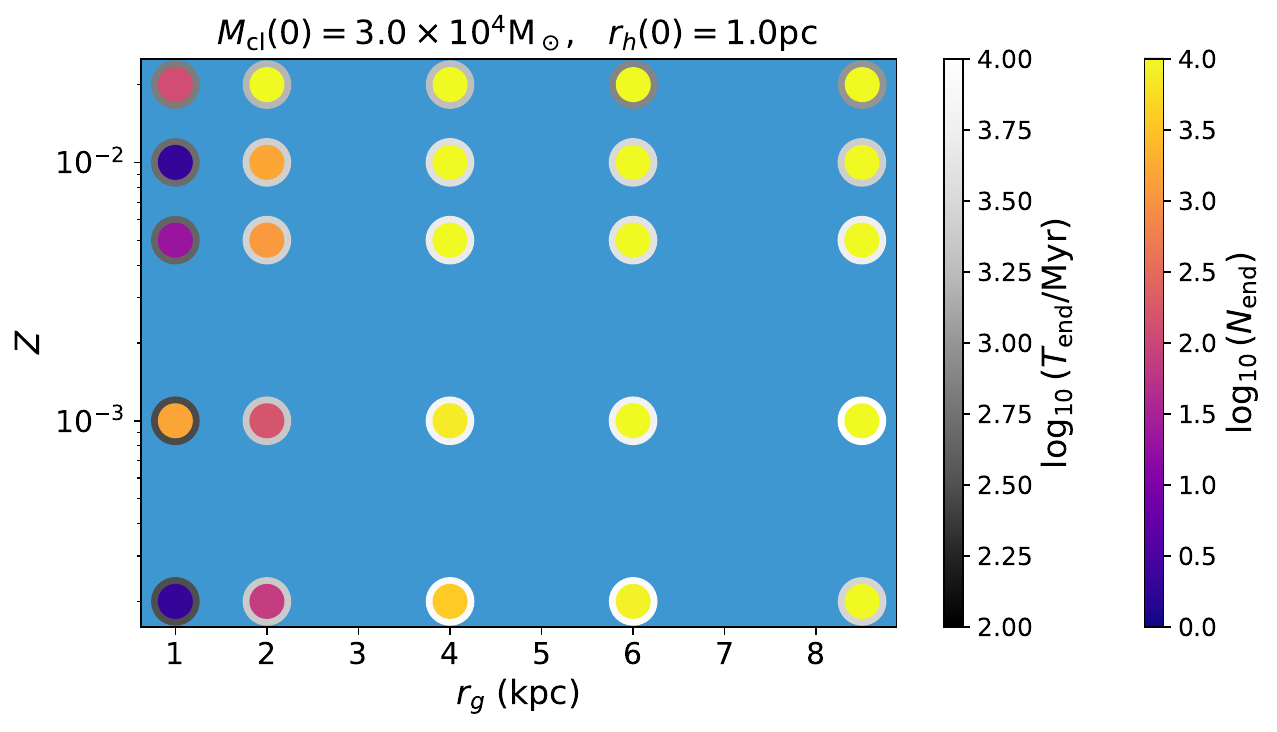}
\includegraphics[width = 8.9 cm, angle=0.0]{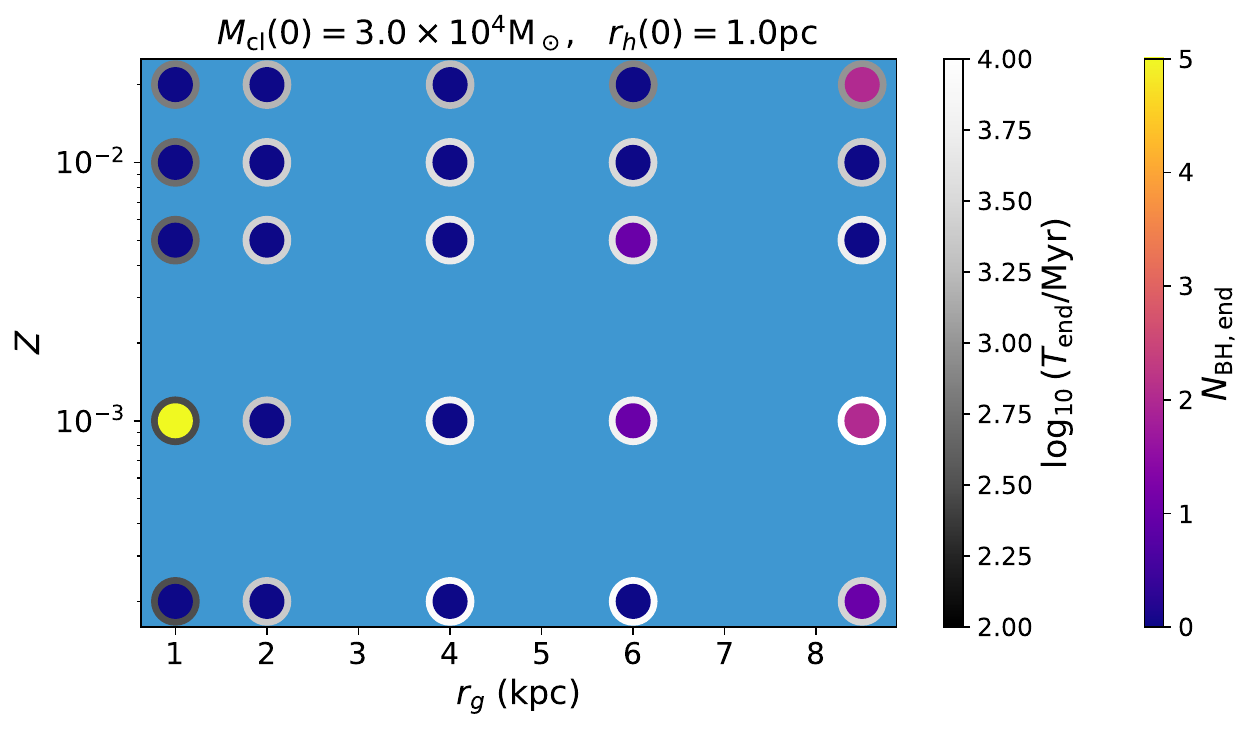}\\
\includegraphics[width = 8.9 cm, angle=0.0]{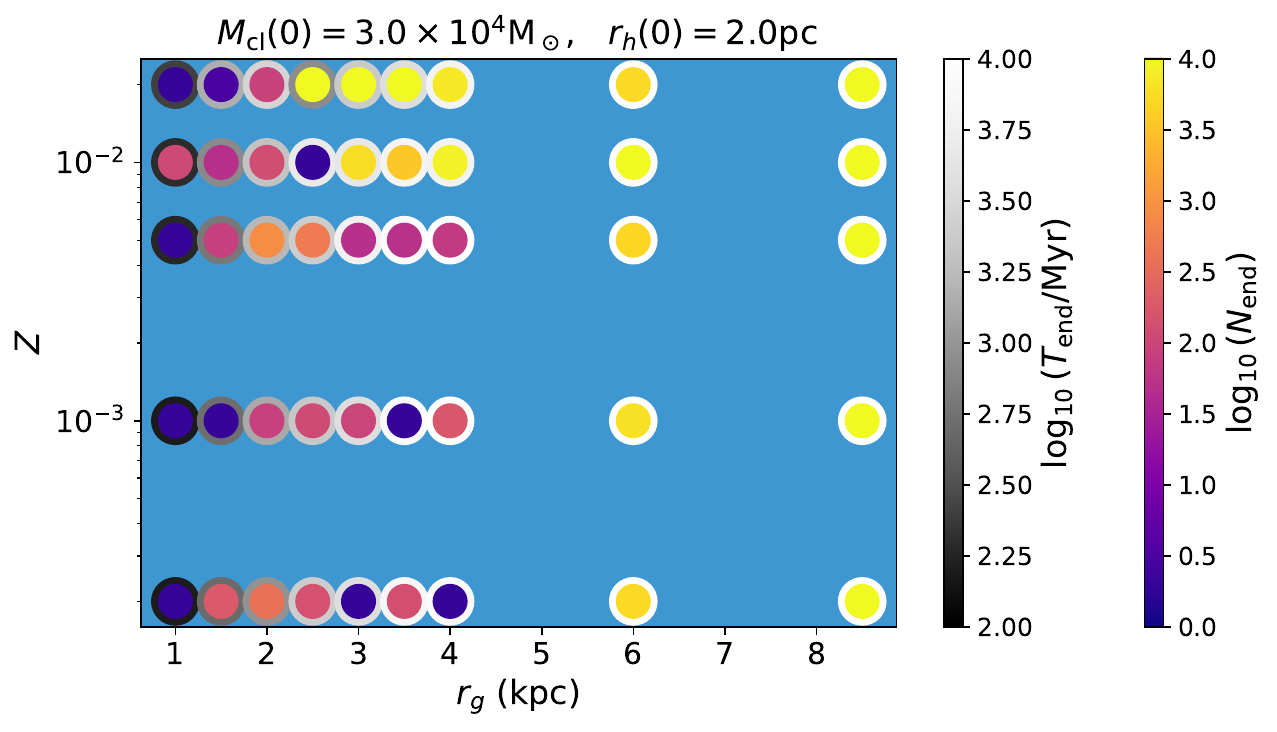}
\includegraphics[width = 8.9 cm, angle=0.0]{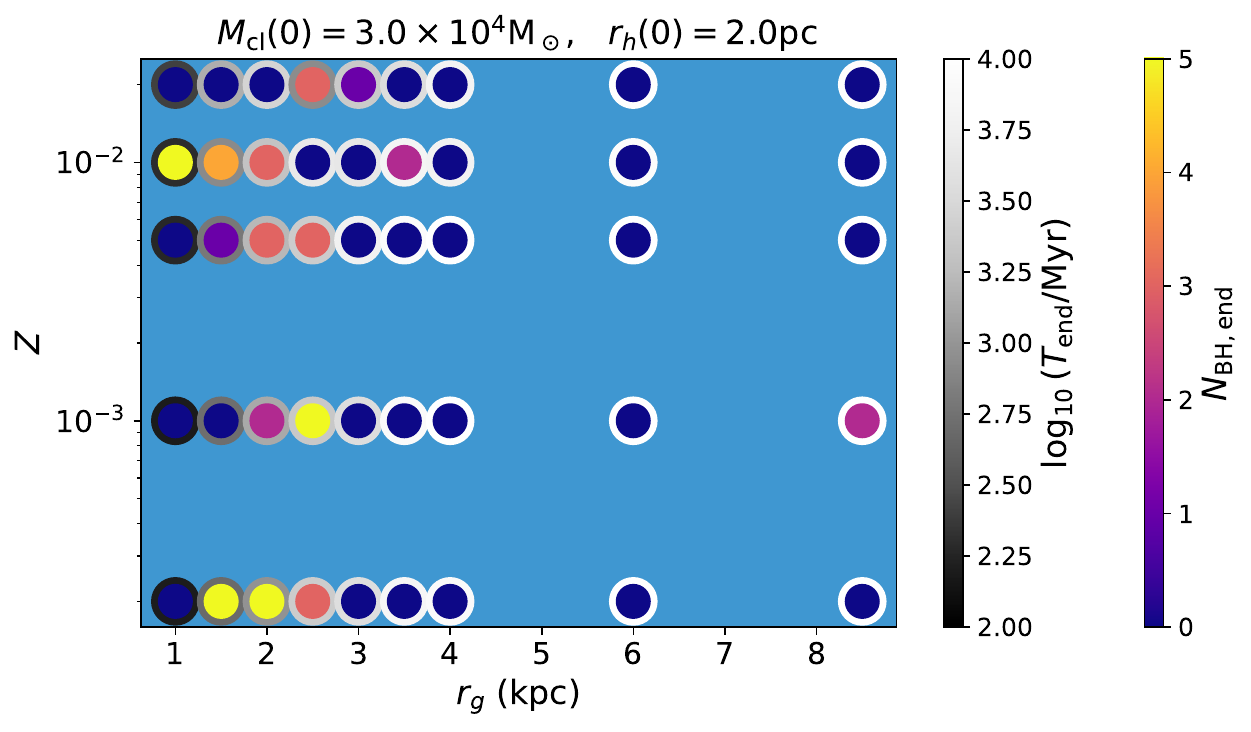}\\
\includegraphics[width = 8.9 cm, angle=0.0]{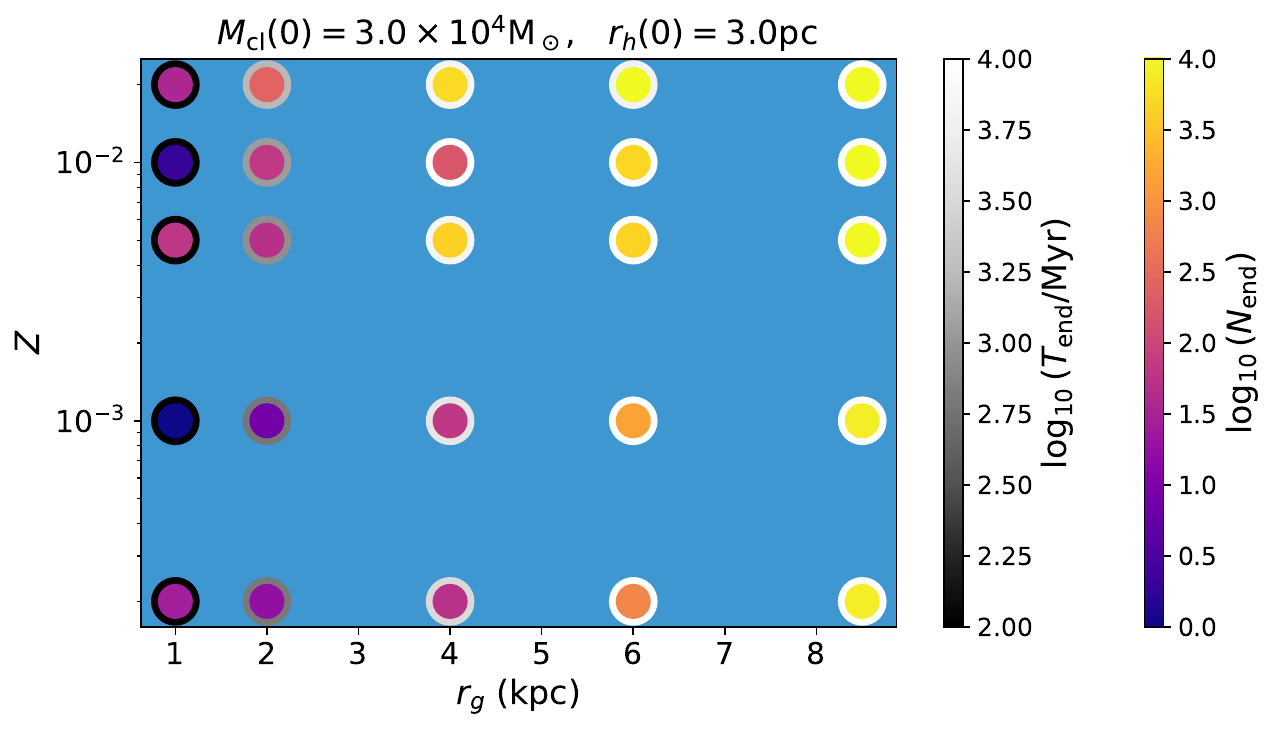}
\includegraphics[width = 8.9 cm, angle=0.0]{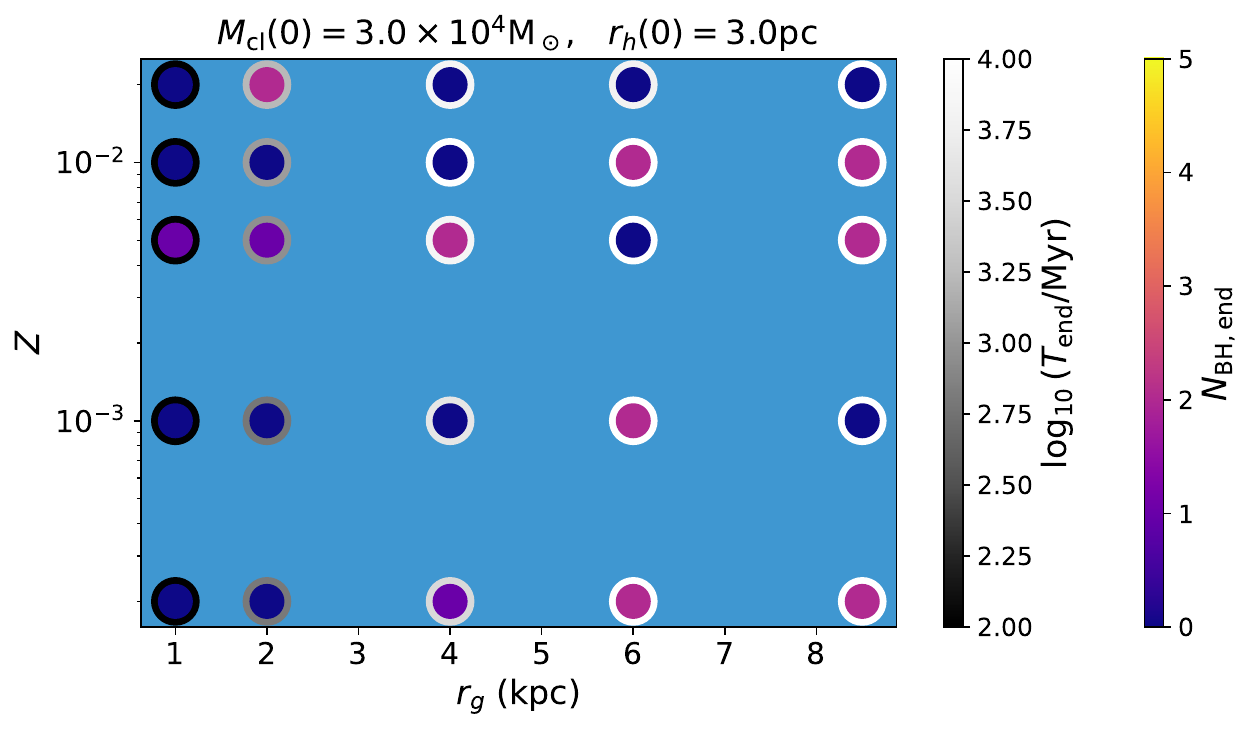}
\caption{Depiction of the computed model grids over galactocentric distance, $r_g$, and
	metallicity, $Z$. The upper, middle,
	and lower panels show the computed grid points (filled circles) for the $\mcl(0)=3.0\times10^4\Ms$
	clusters with $\rh(0)=1$, 2, and 3 pc, respectively. The greyscale color coding
	of the edge of a circle represents the end time in Myr, $\tend$, of the corresponding N-body simulation.
	The fill colour of the circle represents the total number of bound cluster members
	at the end of the simulation, $\nend$ (left panels), and the number
	of BHs bound to the cluster at the end of the simulation, $\nbhend$ (right panels).}
\label{fig:grid}
\end{figure*}

\begin{figure}
\centering
\includegraphics[width = 9.0 cm, angle=0.0]{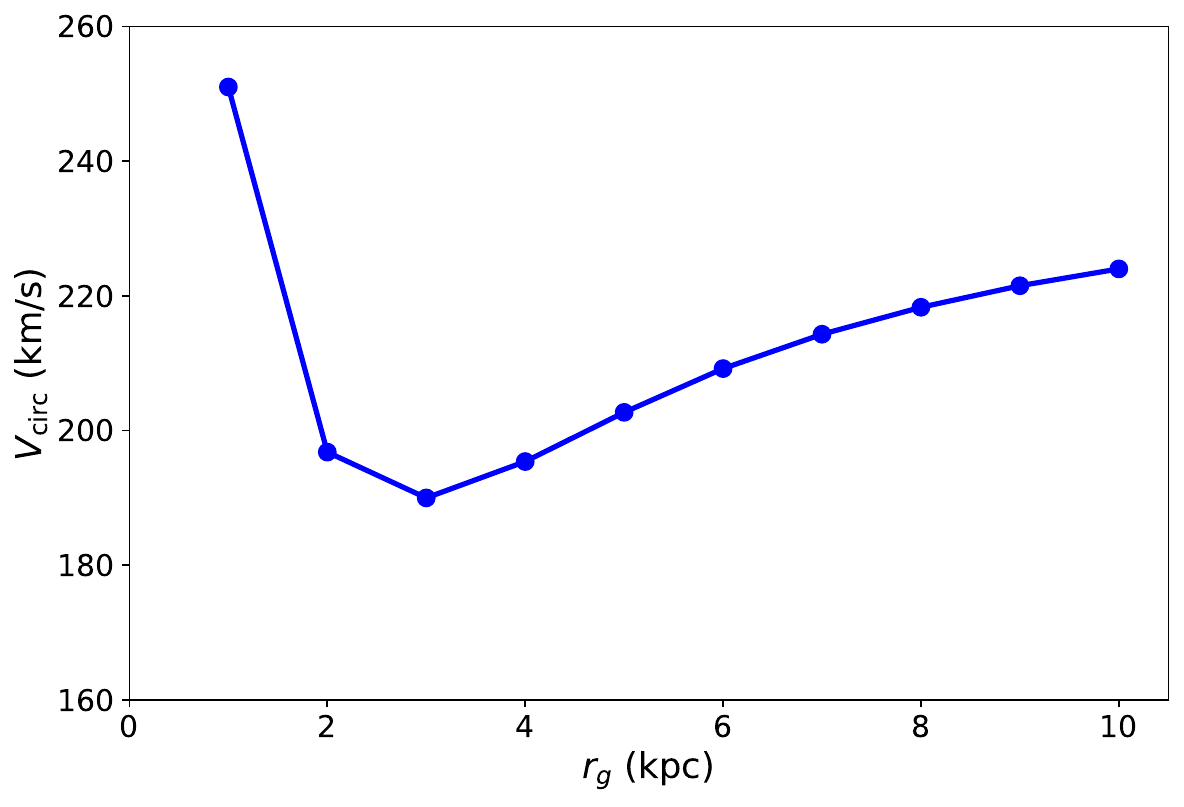}
	\caption{The circular velocity curve (or rotation curve),
	$\vcirc(r_g)$, of the static, axisymmetric galactic gravitational potential
	used in this work. This model field, following that of Ref.~\cite{Allen_1991},
	idealizes the potential due to the Milky Way galaxy.}
\label{fig:vcirc}
\end{figure}

\section{Methods}\label{method}

This section describes the computed grid of evolutionary models of star clusters.

\subsection{Direct N-body simulation of model star clusters}\label{sims}

\subsubsection{The star cluster model grid}\label{grid}

In this study, a homogeneous grid of model star clusters of initial mass $\mcl(0)=3.0\times10^4\Ms$ is evolved.
The initial density and internal velocity dispersion profiles of the clusters are taken to be
according to the Plummer model \citep{Plummer_1911};
which make them initially spherically symmetric and virialized. The initial models have half-mass radii $\rh(0)=1$ pc,
2 pc, and 3 pc. They move on circular orbits on the equatorial plane of an axisymmetric, static galactic potential (see below),
at galactocentric distances $r_g=1.0$ kpc, 2.0 kpc, 4.0 kpc, 6.0 kpc, and 8.5 kpc. Furthermore, at each $r_g$, the
clusters are evolved for the metallicities $Z=0.0002$, 0.001, 0.005, 0.01, and 0.02. For a more detailed exploration
of the impact of the external field, $r_g=1.5$ kpc, 2.5 kpc, 3.0 kpc, and 3.5 kpc are additionally considered for the
$\rh(0)=2$ pc clusters. These choices comprise a grid of a total of 95 evolutionary cluster models spanning
over wide ranges (a factor of few to several orders of magnitude) of $r_g$, $\rh(0)$, and $Z$, 
as depicted in Fig.~\ref{fig:grid}.

In this work, the model clusters are subjected to an axisymmetric, stationary external potential,
that idealizes the gravitational potential of the Milky Way galaxy (hereafter MW). Specifically, 
a central mass-disk-halo model mimicking the MW potential, as in Ref.~\cite{Allen_1991}, is adopted.
This three-component potential is composed of a spherically symmetric potential corresponding to
a central mass of $M_g=1.41\times10^{10}\Ms$, a Miyamoto-Nagai disk potential with
total disk mass $M_d=8.56\times10^{10}\Ms$, radial scale length $A=5.3178$ kpc, and vertical scale length $B=0.2500$ kpc,
and a logarithmic halo potential with a circular velocity of $\vcirc=220.0\kmps$ at a central
distance of $r_g=8.5$ kpc, matching approximately with the solar system.
The potential is generated with the internal galactic potential generator of $\nbseven$ (see below). The rotation
curve corresponding to this model galactic potential is shown in Fig.~\ref{fig:vcirc}.

It is noteworthy that from the perspective of GR mergers in the Universe, MW holds no specific significance.
Given the tediousness associated with direct N-body integrations (see below), it is not practicable to explore
model grids for a range of external potentials. The idealized MW-like potential employed here facilitates
the variation of the external field that the model cluster is subjected to by varying $r_g$.
From the standpoint of the impact of the external field on a cluster's evolution and lifetime, what mainly matters is the
extent of the cluster's initial filling of the tidal lobe. 
The above choice of the external field doubles as a model for MW-specific questions, \eg, cluster-formed
BH-star binaries in MW (see Sec.~\ref{bhms}).

The selection of $\mcl(0)=3.0\times10^4\Ms$ is driven by the fact that the occurrence of GR mergers tends to
become substantial beyond this mass \citep{Banerjee_2020c} and, at the same time, such clusters
are significantly affected by the external field, enabling the study of this effect.
The choice of initial cluster size between 1-3 pc is consistent with the observed sizes of YMCs in
the Milky Way and nearby galaxies \citep{PortegiesZwart_2010}.
The choice of a circular, equatorial cluster orbit is motivated by the fact that clusters formed within
the Galactic gas disk are generally expected to co-rotate with the disk. See Sec.~\ref{summary}
for further discussions.

\begin{figure*}
\centering
\includegraphics[width = 17.5 cm, angle=0.0]{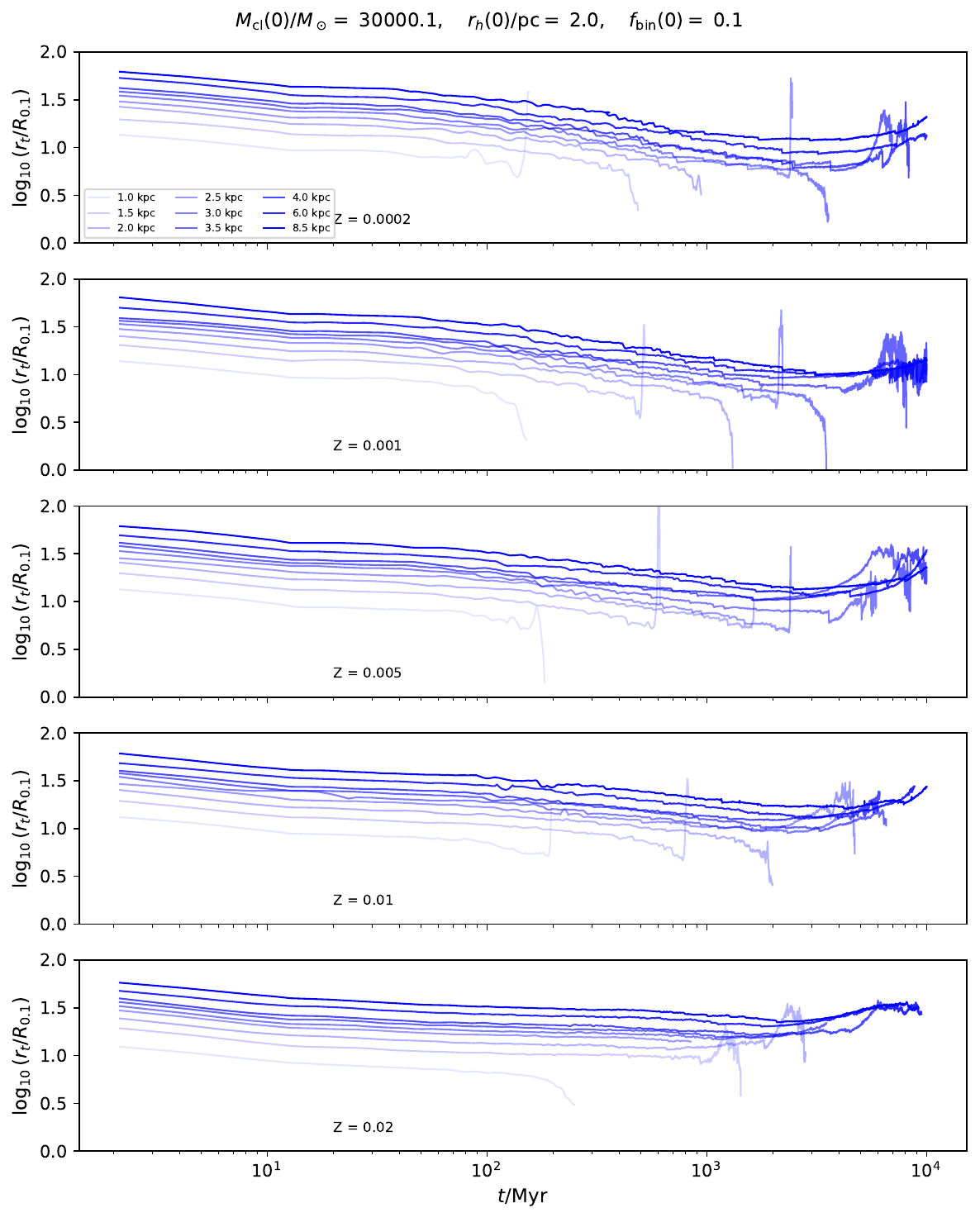}
	\caption{Time evolution of the concentration parameter, as measured by $\log_{10}(r_t/r_{0.1})$,
	for the computed model clusters with $\rh(0)=2.0$ pc (solid lines). The uppermost to
	the lowermost panel plots, in order, the clusters with $Z=0.0002$, 0.001, 0.005, 0.01, and 0.02.
	In each panel, the models corresponding to the different $r_g$s are
	distinguished by varying the shades of the lines (legend).}
\label{fig:conc}
\end{figure*}

\begin{figure*}
\centering
\includegraphics[width = 17.5 cm, angle=0.0]{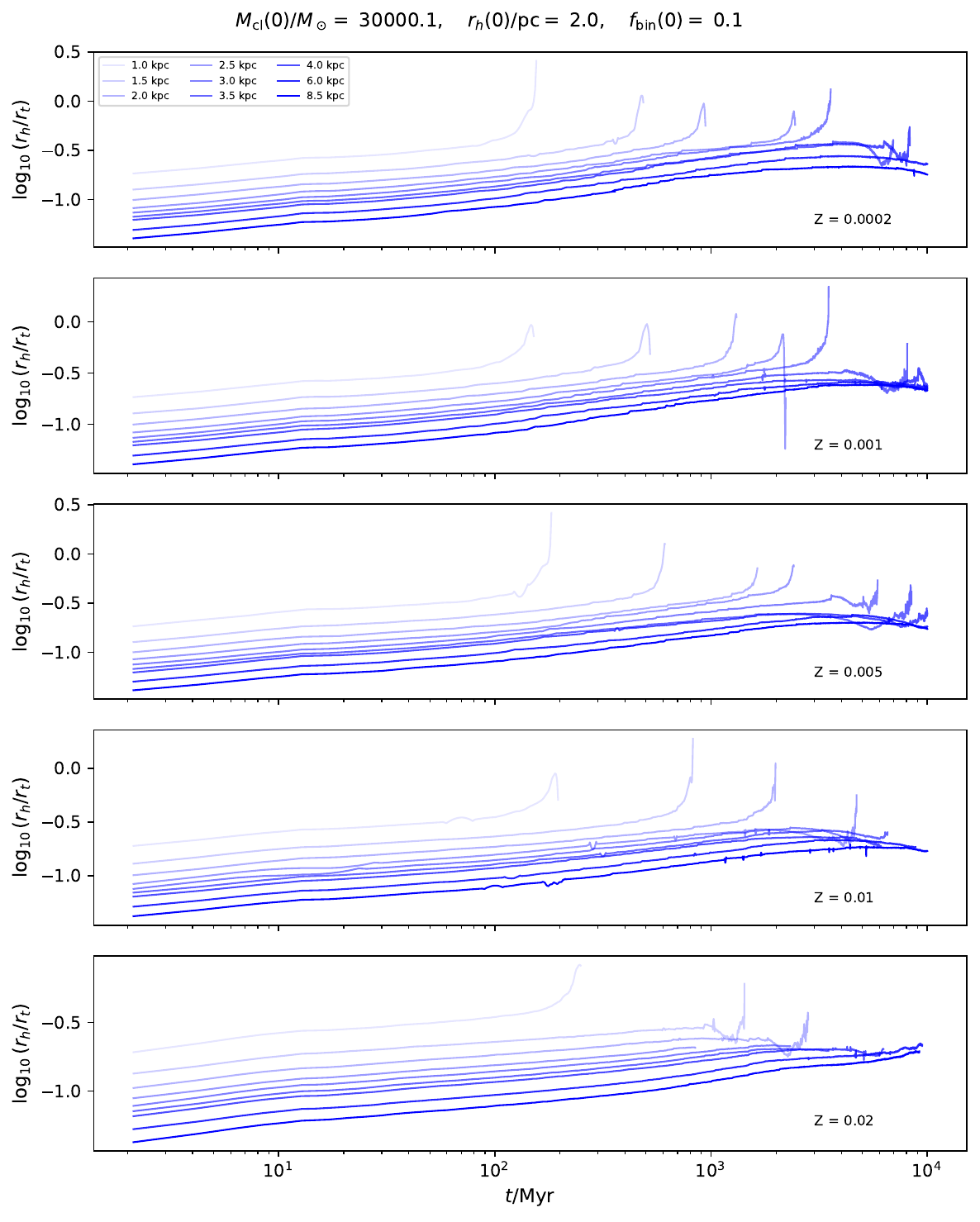}
	\caption{Same as Fig.~\ref{fig:conc}, except that the tidal
	fill factor, $\log_{10}(r_h/r_t)$, is plotted along the Y-axes.}
\label{fig:fill}
\end{figure*}

\subsubsection{Direct, star-by-star, post-Newtonian N-body simulations}\label{nbsims} 

Each model clusters in the above described grid is evolved with the direct, post-Newtonian (hereafter PN)
N-body code $\nbseven$ \citep{Aarseth_2012}. An updated version of $\nbseven$ as detailed in
Refs.~\cite{Banerjee_2020} and \cite{Banerjee_2020c} is used.
These updates implement more recent stellar wind mass loss and remnant-formation prescriptions in the
code, and also implement numerical relativity (hereafter NR)-based GR merger recoil
and spin recycling of BHs that are formed by in-cluster BBH mergers.
The latter implementation allows for runtime, consistent treatment of in-cluster BBH merger sequences;
in particular, the treatment tends to prevent catastrophic mass growth of BHs via
GR mergers.

Runtime stellar and binary evolution is achieved in $\nbseven$ through its coupling with the rapid,
analytical binary evolution code $\bse$ \citep{Hurley_2000,Hurley_2002}, and PN
evolution of compact binaries and higher order systems is computed runtime by incorporating
the few-body integrator $\archain$ \citep{Mikkola_1999,Mikkola_2008}. In the present computations,
the `delayed' remnant-mass model, along with pair-instability supernova (hereafter PSN)
and pulsation-pair-instability supernova (hereafter PPSN) are employed \citep{Fryer_2012,Belczynski_2016a,Banerjee_2020}.
The delayed remnant-mass prescription produces stellar remnants within the so-called `lower mass gap'
between $2\Ms-5\Ms$ -- in other words, no lower mass gap exists for the chosen remnant-mass prescription.
This choice is consistent with the LVK observations of GR mergers involving such lower-mass-gap
compact objects \citep{Unequal_masss_2020,LowerMassGap2024}.

At formation, an NS or a BH is assigned a supernova (hereafter SN) natal kick that is chosen
from a Maxwellian distribution of dispersion $\sigma=265\kmps$, which is then reduced
based on the SN mass fallback fraction as of the prescriptions of Ref.~\cite{Fryer_2012}.
In the present computations, the momentum-conserving fallback speed reduction as in Ref.~\cite{Banerjee_2020}
is applied, resulting in retention at birth of BHs that are $\gtrsim 8\Ms$. All low mass BHs and NSs that are
formed via core-collapse SN escape the cluster at formation. However, NSs formed via
electron capture supernovae (hereafter ECS) receive small natal kicks \citep{Podsiadlowski_2004}
and are typically retained in the cluster at their formation.

The stars in the initial models are taken to be zero age main sequence (hereafter ZAMS) stars 
of masses $0.08\Ms\leq m_\ast\leq150.0\Ms$ that are distributed according
to the Kroupa initial mass function (hereafter IMF) \citep{Kroupa_2001}.
In all initial models, the stars have an overall (see below) primordial-binary fraction of $\fbin=10$\%.  
However, the initial binary fraction of the
O-type stars ($m_\ast\geq16.0\Ms$), which are initially paired only among themselves,
is taken to be $\fobin(0)=100$\%, which is consistent with the observed high binary fraction
among O-stars in young clusters and associations \citep[e.g.][]{Sana_2011,Sana_2013,Moe_2017}.
The O-star binaries are considered to initially follow
the observed orbital-period distribution of Ref.~\cite{Sana_2011}
and a uniform mass-ratio distribution.
The initial orbital periods of the non-O-star primordial binaries follow the period
distribution of Ref.~\cite{Duq_1991} and their mass-ratio distribution is also uniform.
The initial eccentricity of the O-star
binaries follows the eccentricity distribution of Ref.~\cite{Sana_2011} and that
for the rest of the binaries follows the thermal eccentricity distribution
\citep{Spitzer_1987}. As explained in Ref.~\cite{Banerjee_2017b}, such a scheme for including
primordial binaries provides a reasonable compromise between the economy of computing
and consistencies with observations.

In the present computations, small birth spins, as per the `MESA' BH-spin model of Ref.~\cite{Belczynski_2020}, are assigned to
all BHs derived from single stars or from members of non-mass-transferring or non-interacting binaries.
In the event of a BH-star merger
(the formation of a BH Thorne--Zytkow object or BH-TZO \cite{TZ_1975}), 
$\ftz=95$\% of the merging star's mass is assumed to be accreted onto the BH.
In star-star mergers, $\fmrg=20$\% of the secondary's
mass ($\leq10$\% of the total merging stellar mass) is assumed to be lost in the merger process
\citep{Gaburov_2008,Glebbeek_2009}. While the extent of mass accretion onto a BH
and mass loss in star-star mergers is ambiguous, the above choices
favour the formation of massive BHs in moderately dense
environments, especially within the PSN mass gap \citep{Langer_2007,Woosley_2017}.
As demonstrated in Ref.~\cite{Banerjee_2022}, that way the extent of massive BH formation in star clusters
is consistent with the BBH merger rate in the PSN gap as observed by LVK \citep{Abbott_GW190521,Abbott_GWTC3_prop}.
The above assumptions make the present N-body computations consistent with those in Ref.~\cite{Banerjee_2022}.

In the present model grid, all clusters are initially tidally under-filled but the tidal filling
increases (\ie, $r_t$ decreases) with increasing strength of the external field, \ie, with decreasing $r_g$
(see Eqn.~\ref{eq:rt_pm}; below). As a cluster evolves, it initially expands due to BH heating (see Sec.~\ref{intro}),
stellar-evolutionary mass loss in stellar wind, SN and planetary-nebula ejecta, and member loss due to
relaxation-driven evaporation and dynamical ejections \citep[\eg,][]{Banerjee_2010,Aarseth_2012,Kremer_2020}. 
As the scale length (half-mass radius, $\rh$) of the cluster approaches $r_t$, the cluster
starts to become tidally stripped \citep{Giersz_2019}, when its membership, $N$,
and, hence, its relaxation time begin to decrease drastically and the latter approaches the dynamical time \citep{Spitzer_1987}.
At the same time, the cluster deviates significantly from spherical symmetry, being
bulged towards the first and the second Lagrangian points. Due to explicit, star-by-star trajectory
integration without any symmetry or equilibrium assumption, $\nbseven$ (as well other NBODYx-series codes)
can reliably and uniquely handle such an extreme evolutionary condition. Most of the models in the
present grid that have reached such a tidally stripping phase could be evolved until the complete dissolution
($N\lesssim50$).

On the other hand, if a cluster remains significantly tidally under-filled throughout its evolution
then it would undergo a relaxation-driven contraction, once the BH heating is quenched due to
the depletion of the BH core \citep{Banerjee_2010,Aarseth_2012,Kremer_2020}. In the presence  
of a population of primordial binaries as in the present models, the contracting system would
eventually enter another balanced-evolutionary phase (see Sec.~\ref{intro}) where the central
energy generation is dominated by close encounters involving binaries
-- often referred to as the `binary burning' or `core burning' phase \citep{Heggie_2003}.  
While in principle possible,
this evolutionary state is rather tedious to handle with a direct N-body code such as $\nbseven$,  
due to frequent, strongly perturbed hierarchical subsystem formation in the dense cluster core \citep{Spurzem_1996}.
While $\nbseven$ is in principle designed to handle the binary burning phase, the computation
often progresses very slowly and/or experiences frequent crashes. Most models in the present grid that reached
the binary burning phase had to be discontinued. 

All model clusters are evolved at least until most BHs are depleted from the cluster due to
the BHs' dynamical ejection. Most of the $\rh(0)=2$ pc and 3 pc models are evolved 
until complete dissolution of the cluster or until 10 Gyr physical evolution time,
whichever is shorter. Owing to the short relaxation time,
a $\rh(0)=1$ pc cluster needs to dynamically evolve for very long ($\sim10^5$ N-body or Henon time at $r_g\geq4$ kpc),
well past its final core-collapse/core-burning state, until its dissolving phase or the 10 Gyr age is reached.
This would be impractically tedious and time-consuming to compute (see above).
Therefore, most $\rh(0)=1$ pc clusters are evolved only up to the point of BH depletion.
Fig.~\ref{fig:grid} depicts the end state of each of the computed models, in terms of
the number of bound members, $\nend$, and the number of bound BHs, $\nbhend$, at the end time, $\tend$,
of the computation. Some of the $r_g\leq2.5$ kpc, $\rh(0)=2$ pc models show a few retained BHs at the end
of the computation -- these models reached a dynamical-timescale dissolving phase
and the last recorded model snapshot still contained a few BHs.

\section{Results}\label{result}

\subsection{General properties of cluster evolution}\label{evolve}

Fig.~\ref{fig:conc} shows the evolution of the cluster concentration parameter, $\kc$
here defined as the ratio of the tidal radius, $r_t$, to the 10\% Lagrangian
radius, $r_{0.1}$, at a given evolutionary time: $\kc\equiv\log_{10}(r_t/r_{0.1})$.
The $\kc$ evolution is shown for all the 45 $\rh(0)=2$ pc models (Sec.~\ref{grid}).
For convenience, this subset of 45 models, which covers the span of $r_g$ with the highest resolution,
will be presented in the following batched figures, unless stated otherwise.
At all $r_g$ and $Z$, a cluster initially expands due to BH heating, leading to a decline
of its $\kc$. However, as the BH heating is weakened or quenched due to dynamical decay
of the BH-core (Sec.~\ref{intro}), the central region of the cluster contracts
as the cluster approaches its final core collapse, resulting in a rapid increase
in $\kc$. While this qualitative behaviour is common at all $r_g$, the
expansion and the re-collapse occur in shorter timescales at a smaller $r_g$
where the external field is stronger. Fig.~\ref{fig:mcl-rh-t} (middle column) shows the
corresponding $\rh$ evolutions that explicitly demonstrate the accelerated cluster
evolution with decreasing $r_g$. 

A stronger external field (smaller $r_t$; see Sec.~\ref{intro}) causes faster
loss of stars through the tidal boundary as the cluster expands due to BH heating.
This, in turn, shrinks $r_t$ (Eqn.~\ref{eq:rt_gen}) faster, self-accelerating
the cluster's mass loss more effectively, as seen in Fig.~\ref{fig:mcl-rh-t} (right column).
The decaying cluster membership reduces the cluster's relaxation time, making
the course of cluster evolution faster with decreasing $r_g$. A cluster's relaxation
timescale, as measured by the relaxation time at the half-mass radius, is given by \citep{Spitzer_1987}
\begin{equation}
\trh = 0.138\frac{N^{1/2}\rh^{3/2}}{\langle m \rangle^{1/2}G^{1/2}\ln\Lambda},
\label{eq:trh}
\end{equation}
where $N$ is the cluster membership, $\langle m \rangle$ is the cluster's members'
mean mass, $G$ is the gravitational constant, and $\ln\Lambda\approx10$ is the
Coulomb logarithm.\footnote{See Ref.~\cite{Antonini_2020} and references therein for a more detailed
variant of Eqn.~\ref{eq:trh}. The simpler form is used here for the convenience
of discussions.} 

Fig.~\ref{fig:fill} shows the time evolution of the tidal filling factor,
$\kf\equiv\log_{10}(\rh/r_t)$. In the early phase of cluster evolution,
as $\rh$ expands and $\mcl$ (and hence $r_t$) decreases,
$\kf$ increases with time. When the external field is weak, the
effect of the late-time contraction of the cluster (see above) dominates,
causing $\kf$ to decrease again at late times. However, for sufficiently strong external fields ($r_g\lesssim4$ kpc),
the cluster evolution is strongly influenced by the rapid mass loss via tidal stripping. In  
this case, $\kf$ either remains nearly constant or continues to increase   
at late evolutionary times.

In general, at a given $r_g$ the lower is the metallicity the larger is the extent
of expansion of the cluster (see Fig.~\ref{fig:mcl-rh-t}, middle column). This is due to the fact that
lower-$Z$ stars produce more massive BHs, making BH-heating more efficient, as noted
in several earlier studies \citep[\eg,][]{Banerjee_2017,Kremer_2020}. For a given
$Z$, a cluster generally expands to a less extent when it is subjected to a stronger external field (by
placing it at a smaller $r_g$). This is caused by enhanced tidal stripping and decay of the
tidal boundary (\ie, $r_t$) with increasing external field.

\begin{figure*}
\centering
\includegraphics[width = 18.5 cm, angle=0.0]{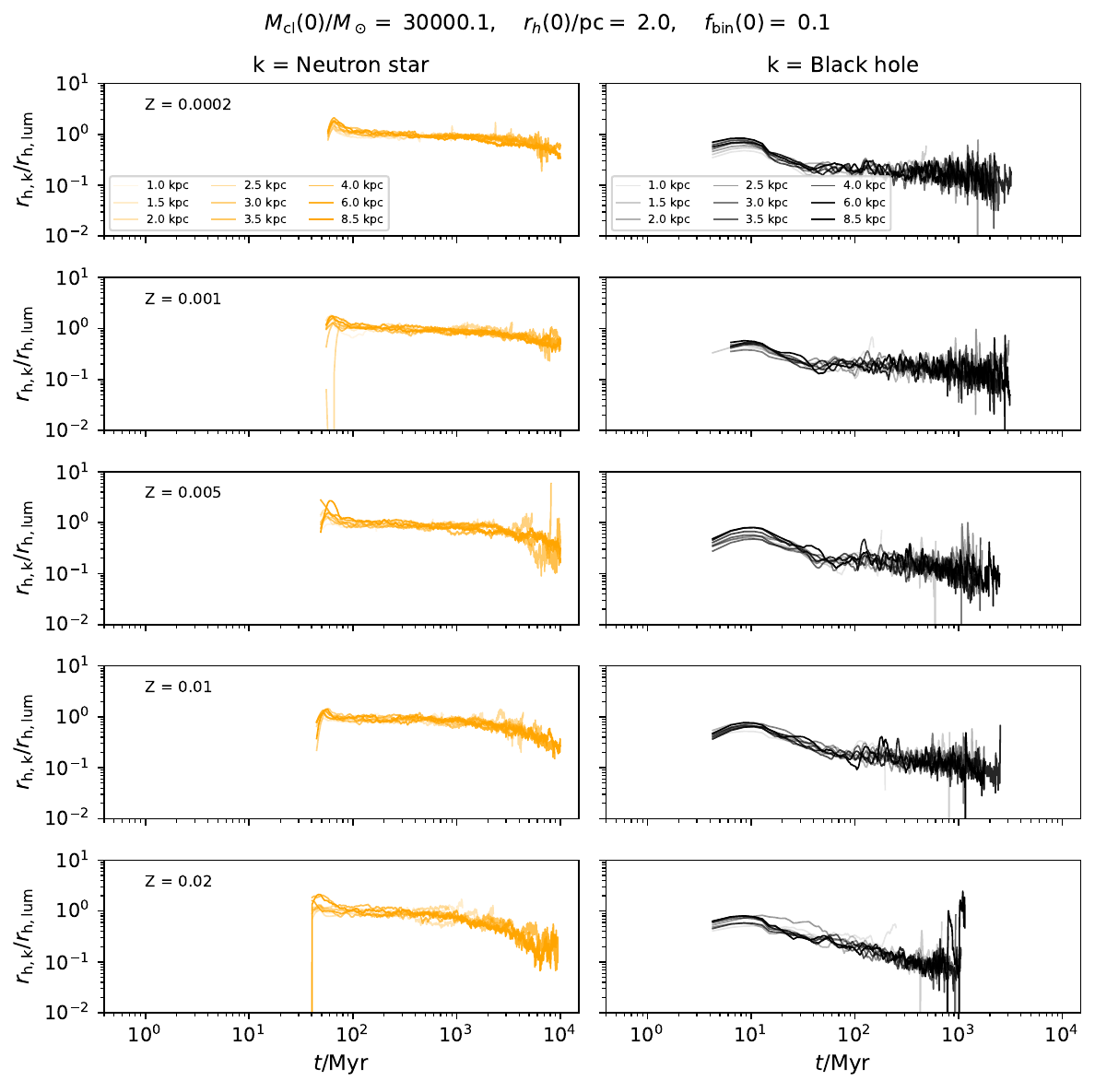}
	\caption{Time evolution of the half-mass radius of the bound NS (left-column panels)
	and BH (right-column panels) members, relative to the half-mass radius of
	the bound luminous members, for the $\rh(0)=2$ pc model clusters (solid lines).
	The panels in the uppermost to the lowermost row
	plot, in order, the clusters with $Z=0.0002$, 0.001, 0.005, 0.01, and 0.02.
	In each panel, the models corresponding to the different $r_g$s are
	distinguished by varying the shades of the lines (legend).}
\label{fig:rhratio}
\end{figure*}

Fig.~\ref{fig:rhratio} shows the ratio of the half mass radii of the in-cluster NS (left column)
and BH (right column) subpopulations, denoted by, respectively, $\rhns$ and $\rhbh$, to the
half mass radius of the luminous members, $\rhlum$, for the $\rh(0)=2$ pc models.
A key feature that this figure demonstrates is that the efficient mass segregation of BHs and BH
core formation (Sec.~\ref{intro}) take place irrespective of the strength of the external field
and metallicity. While a stronger tidal field causes to dissolve the cluster faster it also
accelerates the mass segregation process (Sec.~\ref{intro}). Due to the BH heating effect,
NSs begin to segregate towards the cluster center only when the BH core is sufficiently
depleted, as noted in earlier studies \citep{Banerjee_2017b,Ye_2019}.

Due to the retained BH population being centrally segregated, the clusters generally become increasingly remnant rich
as they evolve. This is demonstrated in Fig.~\ref{fig:darkfrac}, which shows that a stronger tidal
stripping causes the remnant mass fraction to increase at a faster rate and makes
the cluster increasingly dominated by dark remnants (BHs and NSs). This effect is qualitatively
similar to the findings from other studies (\eg, \cite{Banerjee_2011,Giersz_2019}).
\footnote{The lines in Figs.~\ref{fig:conc}, \ref{fig:fill}, \ref{fig:rhratio}, \ref{fig:mcl-rh-t},
and \ref{fig:darkfrac} are moderately smoothed by applying the Savitzky-Golay-Filter
(with 10 points and polynomial order 3), as available in the {\tt SciPy} module
{\tt scipy.signal.savgol\_filter}.}

\begin{figure*}
\centering
\includegraphics[width = 14.5 cm, angle=0.0]{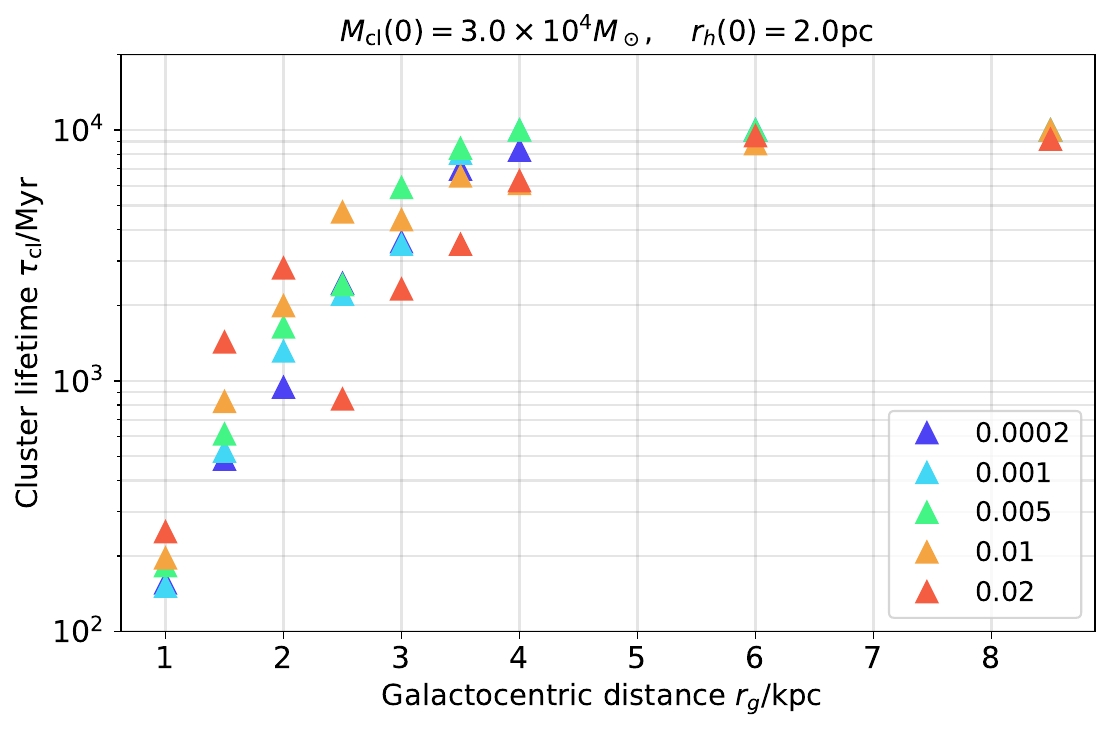}
	\caption{Cluster lifetime, $\taucl$ (Y-axis), as a function of $r_g$ (X-axis)
	and $Z$ (legend), for the $\rh(0)=2$ pc model clusters (filled triangles).
	For $r_g>4$ pc, the clusters have not dissolved by the maximum
	cluster evolution time, $10^4$ Myr. For these models,
	$\tend\approx10^4$ Myr are plotted along the Y-axis,
	representing lower limits of $\taucl$.
	The $Z=0.02$ models, placed within $2.5{\rm~pc}\leq r_g \leq4.0{\rm~pc}$,
	could not be evolved until dissolution (see Fig.~\ref{fig:grid}, text); the simulation end
	times are plotted as the lower limits of $\taucl$ for these models too.}
\label{fig:taucl}
\end{figure*}

Fig.~\ref{fig:taucl} shows the cluster lifetime, $\taucl$, of all the $\rh(0)=2$ pc models.
Here, $\taucl$ is defined as the minimum between the evolution time until $N\leq50$ members are left bound
or $10^4$ Myr; see the caption of Fig.~\ref{fig:taucl} for further details. As seen in the figure,
$\taucl$ overall decreases with decreasing $r_g$, \ie, increasing external field, as expected.
On the other hand, $\taucl$ increases with metallicity. Stars at higher metallicity produces
less massive BHs due to stronger wind mass loss \citep{Belczynski_2010}, weakening the BH heating and the
cluster expansion that it drives. This, in turn, causes less stripping of the cluster through
the tidal boundary, \ie, diminishes the cluster's tidal dissolution rate. Furthermore, at higher
metallicity, stellar lifetime increases, causing the mass loss associated with the
formation of stellar end state (comprising mainly SN and planetary-nebula mass loss)
to occur at a lower rate, also resulting in weaker expansion and dissolution of the cluster.  

\begin{figure*}
\centering
\includegraphics[width = 8.9 cm, angle=0.0]{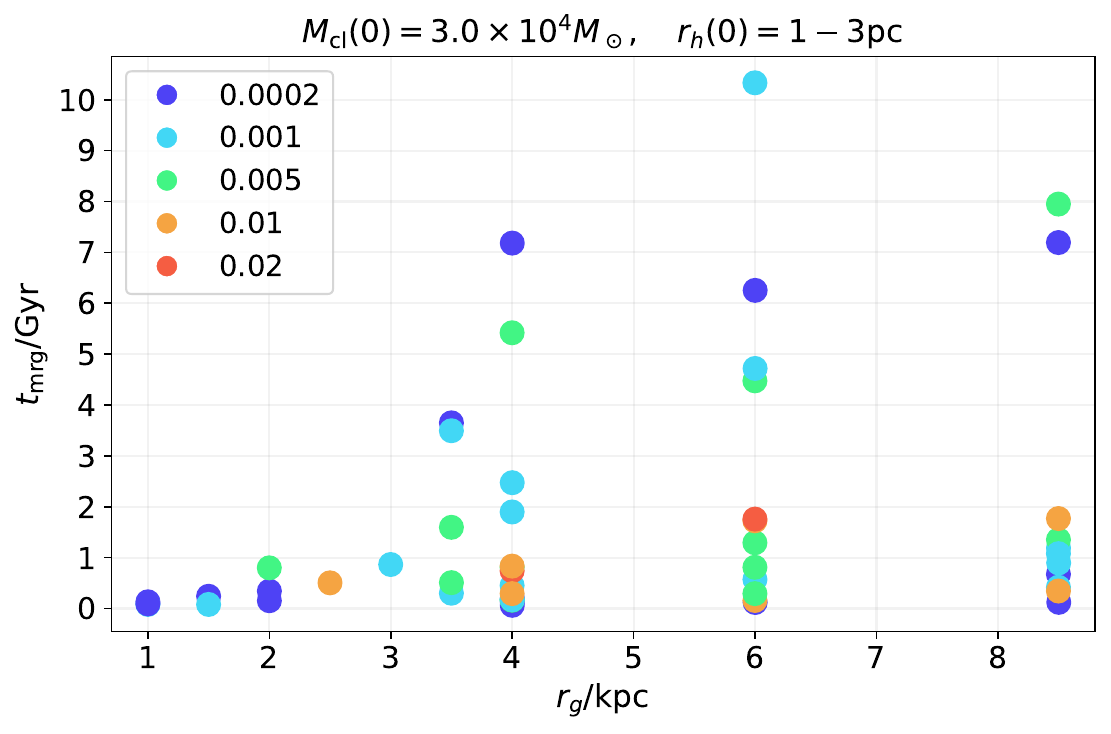}
\includegraphics[width = 8.9 cm, angle=0.0]{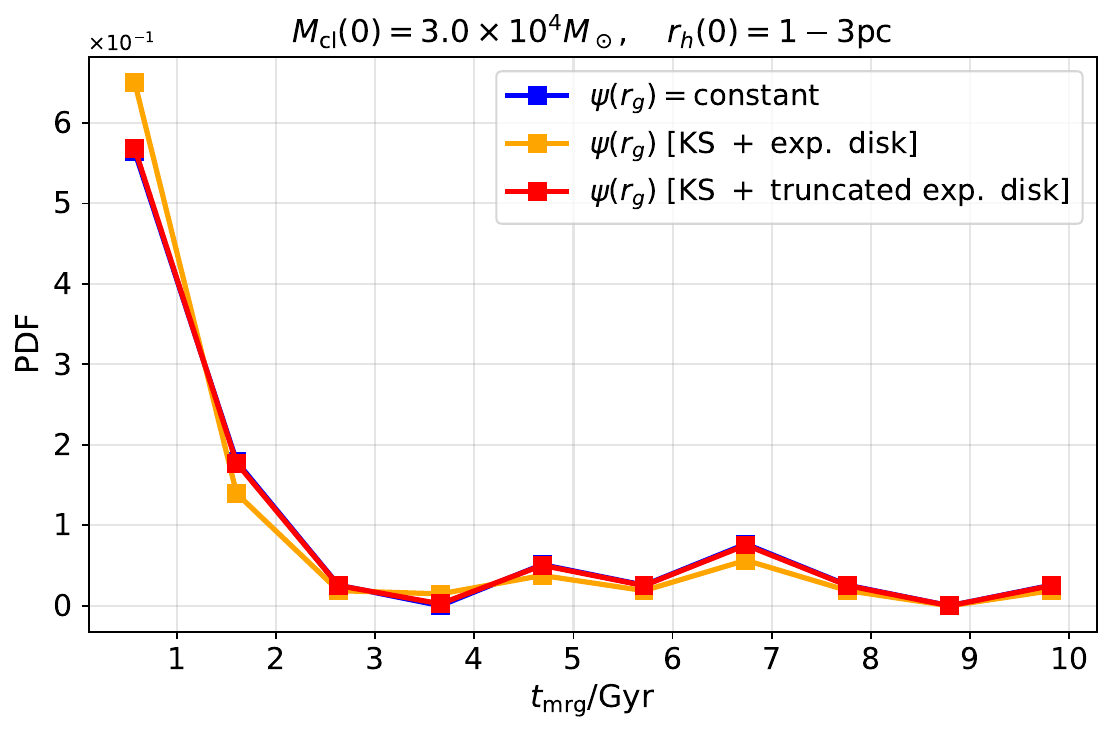}
	\caption{{\bf Left panel:} delay time of GR mergers, $\tmrg$ (Y-axis), as a function
	of $r_g$ (X-axis) and $Z$ (legend), for all the 95 computed model clusters
	in this study (filled circles).
	{\bf Right panel:} the corresponding normalized distributions (or PDFs) of $\tmrg$, for different
	galactic SFR profiles (legend).}
\label{fig:tmrg}
\end{figure*}

\subsection{Formation of gravitational-wave mergers}\label{gwmrg}

Due to the formation of a strongly mass-segregated BH sub-population or BH-core (Sec.~\ref{evolve}),
the model clusters have the potential to form dynamical BBHs and their GW mergers (Sec.~\ref{intro}).
The majority of the present computed models have produced one or more BBH mergers, and they comprise
a combination of in-cluster and ejected mergers. However, due to the inefficient mass
segregation of the NS-members (see Fig.~\ref{fig:rhratio} and the associated
references), no GW merger involving an NS has occurred in these models.
Notably, a recent study \citep{Barber_2025} suggests that the presence of a high fraction of primordial massive
binaries in clusters, as in the current models, leads to BBH mergers that are, in majority,
derived from the isolated evolution of the primordial binaries (\ie, as if each of the binaries
has evolved independently, being outside of the cluster). However, this result is in contrast
with previous cluster-evolution studies based on direct N-body
and Monte Carlo methods (\eg, \citep{Banerjee_2017b,Kremer_2020,ArcaSedda_2024c}),
where the model clusters also contained a high fraction of primordial massive binaries of similar properties.
In these works, the majority of the GW mergers
from the clusters have been found to be purely dynamically paired (\ie, where the merging
members' progenitor stars did not belong to the same primordial binary) and/or to be influenced by
dynamical interactions.

For the present set of models, all GW mergers are BBH mergers, 80\%
of which are in-cluster mergers. Practically all (98\%) of the in-cluster mergers are dynamically
paired. In contrast, most (91\%) of the ejected mergers are primordially paired (\ie, where the merging
members' progenitor stars belonged to the same primordial binary).
This is consistent with previous studies \citep{Banerjee_2017,Banerjee_2017b,DiCarlo_2019,Kumamoto_2019,Rastello_2021}
that suggest that typical YMCs and OCs tend to predominantly produce in-cluster,
dynamically paired BBH mergers, owing to their low velocity dispersions and
escape velocities.
A detailed, collaborative study on dynamically paired versus primordially paired  
GW mergers from clusters is currently in progress.

Fig.~\ref{fig:tmrg} (left panel) shows the delay time\footnote{In this study, the delay time of a GW merger is defined
as the time at which the coalescence occurs since the beginning of the parent cluster's evolutionary simulation.},
$\tmrg$, of
all the individual BBH mergers as obtained from the current evolutionary model grid (all 95 models).
They are plotted against their
respective parent cluster's $r_g$, and are distinguished based on the cluster's metallicity. Owing to the 
accelerated cluster evolution (Sec.~\ref{intro}) and shortening cluster lifetime (Sec.~\ref{evolve}; Fig.~\ref{fig:taucl})
with the strengthening external field, the timespan over which the mergers are produced
(\ie, the range of $\tmrg$) shortens with decreasing $r_g$. However, even the clusters that are subjected to the strongest
external field ($r_g=1$ kpc), with $\taucl \sim 100$ Myr, can also produce BBH mergers at the lowest metallicities.

Fig.~\ref{fig:tmrg} (right panel) shows the $\tmrg$ distribution corresponding to a population of clusters, constructed
by sampling the model clusters according to various radial star formation rate (hereafter SFR) profiles, $\sfr$.
In this simple Monte Carlo sampling based on $r_g$, all cluster $\rh(0)$s are taken to be equally likely. Apart
from the uniform SFR, \ie, $\sfr = {\rm ~ constant}$, a Kennicutt-Schmidt-type (hereafter KS) SFR profile
\citep{Kennicutt_1989} modulated by the gas-disk rotation curve \citep{Boissier_2007,Yin_2009} is also applied:
\begin{equation}
	\sfr = \alpha\sgas(r_g)^{1.5}\left(\frac{r_\odot}{r_g}\right).
\label{eq:kennsh}
\end{equation}
Here $\sgas(r_g)$ is the gas surface density profile, and $\alpha=0.1$ and $r_\odot=8.5$ kpc, as appropriate for the MW.
The $\sgas$ profile is taken to be exponential, \ie,
\begin{equation}
	\sgas(r_g) \propto e^{-(r_g-r_\odot)/l_s},
\label{eq:expr}
\end{equation}
with the scale length $l_s=2.3$ kpc, as appropriate for the MW. Additionally, a truncated exponential profile
is also considered, \ie, 
\begin{equation}
	\sgas(r_g) = \begin{cases}
		        {\rm ~constant}                & \text{$r_g\leq r_0$}\\
			\propto e^{-(r_g-r_\odot)/l_s} & \text{$r_g>r_0$},
		       \end{cases}
\label{eq:exprt}
\end{equation}
with the truncation length $r_0=12$ kpc, as reminiscent of the MW \citep{Kalberla_2009}.  
As seen in Fig.~\ref{fig:tmrg} (right panel), the $\tmrg$ distribution is rather insensitive
to the chosen $r_g$ profile, except for the slight bias towards small $\tmrg$ for the
purely exponential radial density profile. The latter feature is due to the stronger
representation of the shorter-lived clusters at smaller $r_g$ that produces smaller-$\tmrg$
mergers (\cf, Fig.~\ref{fig:tmrg}; left panel), for the purely exponential case.
The overall significant bias towards $\tmrg\lesssim2$ Gyr is due to the fact that
these moderately massive model clusters produce most of their GW mergers within this time
frame irrespective of $r_g$ (or the extent of the tidal field). At late evolutionary ages,
the BH engine weakens (Sec.~\ref{intro}) making the cluster inefficient in producing
BBH mergers even if it survives the tidal stripping. Note that the
demonstration in Fig.~\ref{fig:tmrg} (right panel) is specific to a MW-like galaxy, but
it shows that the exact radial distribution of the tidally dissolving clusters within the host galaxy's
potential can generally be considered to be a non-critical factor for the GW merger delay time distribution.

\begin{figure*}
\centering
\includegraphics[width = 18.2 cm, angle=0.0]{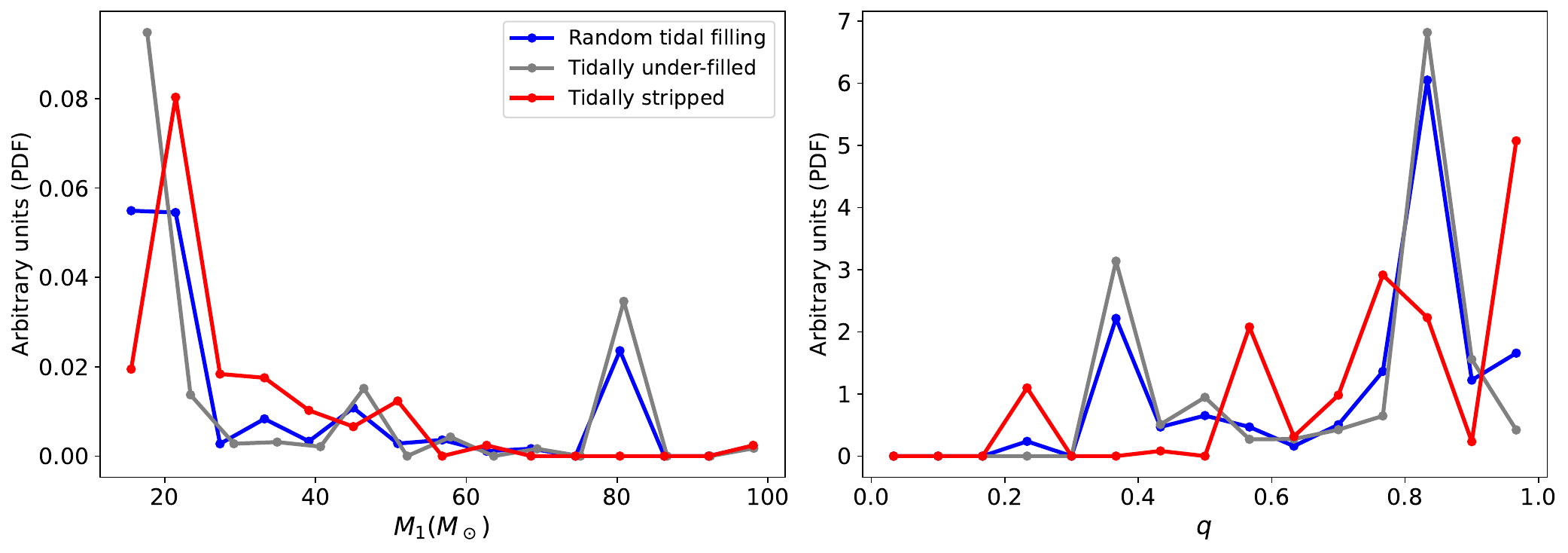}
\includegraphics[width = 12.5 cm, angle=0.0]{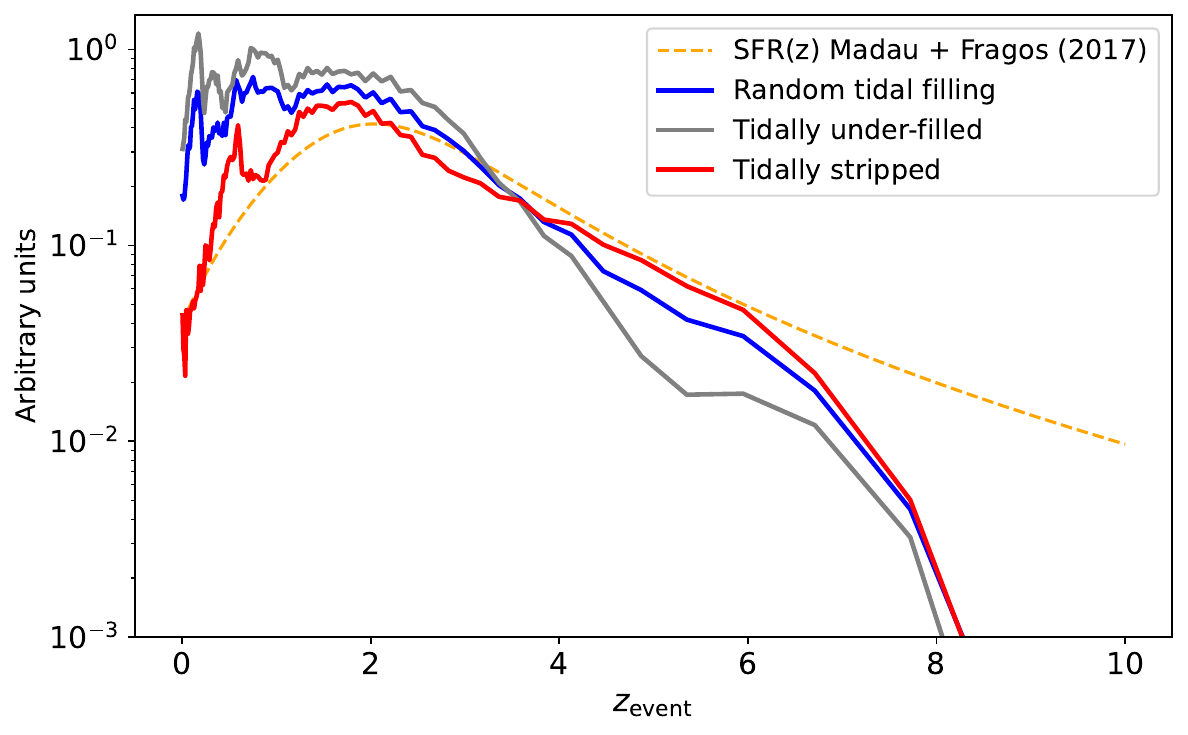}
	\caption{{\bf Upper panels:} present-day-observable, intrinsic distributions of GR-mergers' primary mass, $\mone$ (left),
	and mass ratio, $q$ (right), as obtained from the entire model cluster grid. Three different cases of
	the parent clusters' initial tidal radius filling are compared, as indicated in the legend.  
        All distributions shown in these panels are individually normalized.
	{\bf Lower panel:} distribution of merger redshift, $\zevnt$, for the different tidal radius filling cases
	as indicated in the legend (solid lines).
	The heights of the curves are scaled to be proportional to the integrated number of merger events
	within $\zevnt\leq10$, for the respective tidal-radius-filling cases. For comparison, the profile of
	the considered cosmic star formation rate evolution is over-plotted (dashed line).}
\label{fig:m1dist}
\end{figure*}

\begin{figure*}
\centering
\includegraphics[width = 14.5 cm, angle=0.0]{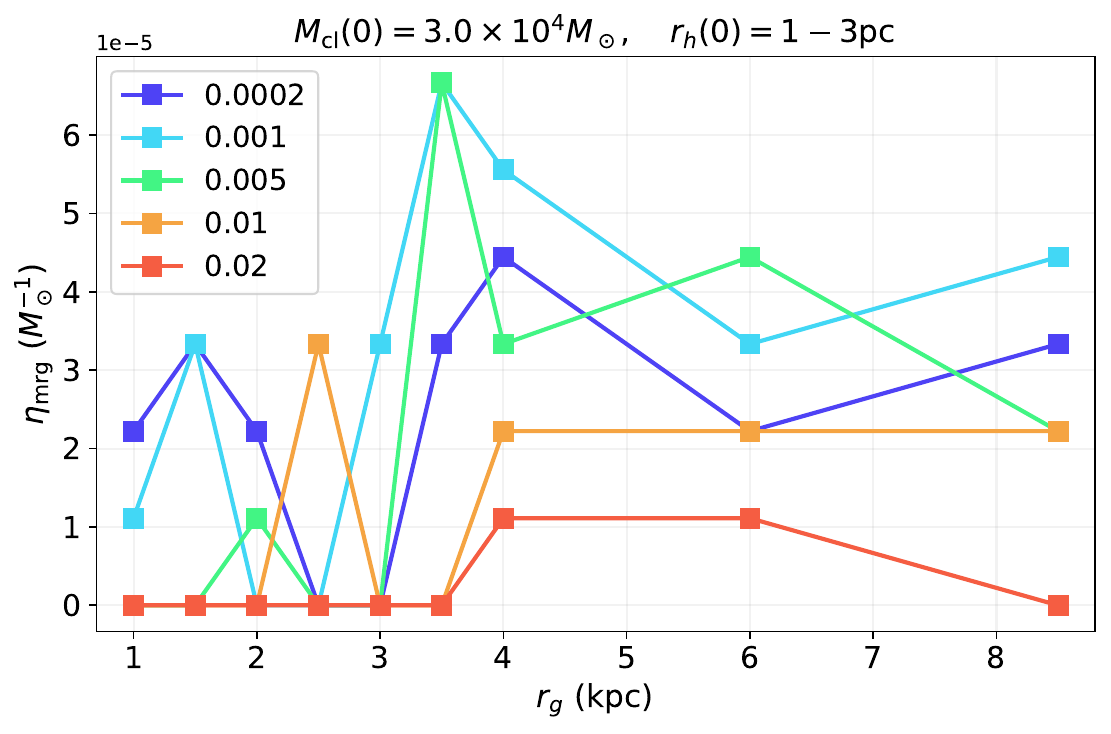}
	\caption{Efficiency of BBH GR mergers per unit cluster mass, $\etamrg$ (Y-axis), as a function of $r_g$ (X-axis)
	and $Z$ (legend), from all the 95 computed model clusters in this study (connected filled squares). See
	text for the details.}
\label{fig:etamrg}
\end{figure*}

Fig.~\ref{fig:m1dist} (upper panels) shows the present-day-observable,
intrinsic distributions of the merger primary mass, $\mone$,
and mass ratio, $q$, as obtained from the entire model grid.
Here, three cases are considered, namely, mergers from (a) only the
clusters at $r_g > 4$ kpc that remain tidally under-filled throughout their evolution and
survive the tidal dissolution (Sec.~\ref{evolve}; Fig.~\ref{fig:taucl}), (b) only the clusters at $r_g\leq4$ kpc 
that fill their tidal radii in the course of their evolution and dissolve in $<10$ Gyr 
due to tidal stripping (the tidally stripped case), and (c) clusters with
$r_g$ chosen randomly within $1.0{\rm ~kpc}\leq r_g \leq 8.5{\rm ~kpc}$ (the random tidal filling case).  
Over the respective $r_g$-ranges, the cluster $r_g$s and $\rh(0)$s are chosen randomly.  
The present-day-observable distributions are obtained by applying the Monte Carlo approach as 
detailed in Ref.~\cite{Banerjee_2021b}, which description, for brevity, is not repeated
here. As in the above reference, the cosmic SFR evolution as of Ref.~\cite{Madau_2017} (up to redshift $z=10$),
the redshift-metallicity dependence as of Ref.~\cite{Chruslinska_2019} (their `moderate-$Z$' dependence), 
a GW event visibility horizon of $\zmax=1$, and the $\Lambda$CDM cosmological framework
\citep{Peebles_1993,Narlikar_2002,Wright_2006} with the cosmological constants from Planck
($H_0=67.4\kmps{\rm~Mpc}^{-1}$, $\Omega_{\rm~m}=0.315$, flat Universe
for which $\thub=13.79{\rm~Gyr}$ \cite{Planck_2018}) are adopted.
A total of $10^6$ clusters are drawn from the model grid for the Monte
Carlo sampling \citep{Banerjee_2021b}, in each case. Since, in this study, only one cluster mass
is considered, only the normalized distributions (probability density function or PDF) are plotted
in Fig.~\ref{fig:m1dist} (upper panels) and not the total (differential) merger rate distributions. That way,  
Fig.~\ref{fig:m1dist} (upper panels) shows only the forms of the distributions in the different
cases and the differences between them. As seen, the $\mone$ and $q$ distributions are overall similar 
in form for the three cases. A few mergers occur within the PSN gap (\ie, with $\mone>40\Ms$),
mainly due to the formation of over-massive BHs via star-star mergers
\citep{Spera_2019,Banerjee_2020,DiCarlo_2020b,Banerjee_2022}.

Fig.~\ref{fig:m1dist} (lower panel) shows the distributions of merger redshift, $\zevnt$,
for the above-mentioned cases. The redshift bins are taken such that they correspond to
equal intervals of cosmic age, \ie, the curves represent the form of the merger rate density
evolution. The curves are normalized such that their integrals
are proportional to the total number of events within $\zevnt\leq10$ (for
$3\times10^5$ draws from the model grid in each case). The redshift evolution and the magnitude
of the merger rate density are similar in all the three cases for $\zevnt\gtrsim1.6$, and they all peak
at $\zevnt\approx1.6$, \ie, at a somewhat later epoch than that of the SFR peak (at redshift 2.1).
The forms of the merger rate evolution generally do not follow the SFR evolution:
across $\zevnt\approx1.6$, they decline faster than the SFR with increasing redshift
and decay slower than the SFR with decreasing redshift.
However, for $\zevnt\lesssim1.6$, the merger rate density evolution from the tidally stripped population tends to follow
the SFR and drops off faster than the other cases.

These properties follow from
the fact that the delay time of BBH mergers from the $r_g>4$ kpc models (that are included in the
under-filling and the random-filling cases) ranges from 100s of Myr to several Gyr,
and the metal poor clusters ($Z\leq0.005$) exclusively produce the very long delay time
mergers with $\tmrg>2$ Gyr (Fig.~\ref{fig:tmrg}, left panel). Since
the high-redshift clusters are metal poor due to the adopted metallicity-redshift dependence (see above),
the long delay time mergers from them offset the merger rate evolution towards
low redshifts (late epochs).
On the other hand, for the tidally stripped clusters ($r_g\leq4$ kpc),
$\tmrg<1$ Gyr for most mergers -- long-term mergers barely take place in these models. 
This causes the mergers from the stripped clusters to follow the SFR evolution
pattern more closely, especially at later (lower) redshifts when the change of redshift with
cosmic age is smaller \citep{Peebles_1993,Narlikar_2002}. Due to the use of a single
$\mcl(0)$ in this study, no comparison with the LVK merger rate density is presented here.
Such comparisons have been presented in detail in earlier studies by the author (\eg, \cite{Banerjee_2021b}),
and further works in such lines are in preparation.

Fig.~\ref{fig:etamrg} plots the BBH merger efficiency, $\etamrg$, as a function of $r_g$, for
different metallicities. Here, the merger efficiency of a population of clusters with total
initial mass $M_{\rm cl,tot}(0)$ is defined as
\begin{equation}
\etamrg \equiv \frac{N_{\rm mrg,tot}}{M_{\rm cl,tot}(0)},
\label{eq:etamrg}
\end{equation}
where $N_{\rm mrg,tot}$ is the total number of GR merger events produced by the cluster population.
Due to the formation of more massive BHs and larger BH retention fraction (\eg, \cite{Banerjee_2020}), the merger
efficiencies are generally higher at lower metallicities, in agreement with previous studies
(\eg, \cite{Rastello_2021}). Additionally, a $r_g$-dependence of $\etamrg$ is also apparent from
Fig.~\ref{fig:etamrg}. For most metallicities, $\etamrg$ tends to initially increase with decreasing $r_g$,
thereafter decline again as $r_g$ decreases further. The $\etamrg$ maximizes around $r_g\approx4$ kpc,
where the cluster begins to become tidally stripped (Sec.~\ref{evolve}). This behaviour of $\etamrg$ is a manifestation
of the counteracting effects of depletion and accelerated dynamics in tidally stripping star clusters (Sec.~\ref{intro}).
This peaked behaviour of $\etamrg$ also causes the non-stripped, partially stripped, and predominantly stripped  
cluster populations to produce BBH mergers at comparable rates (Fig.~\ref{fig:m1dist}; lower panel).

\begin{figure*}
\centering
\includegraphics[width = 0.99\linewidth, angle=0.0]{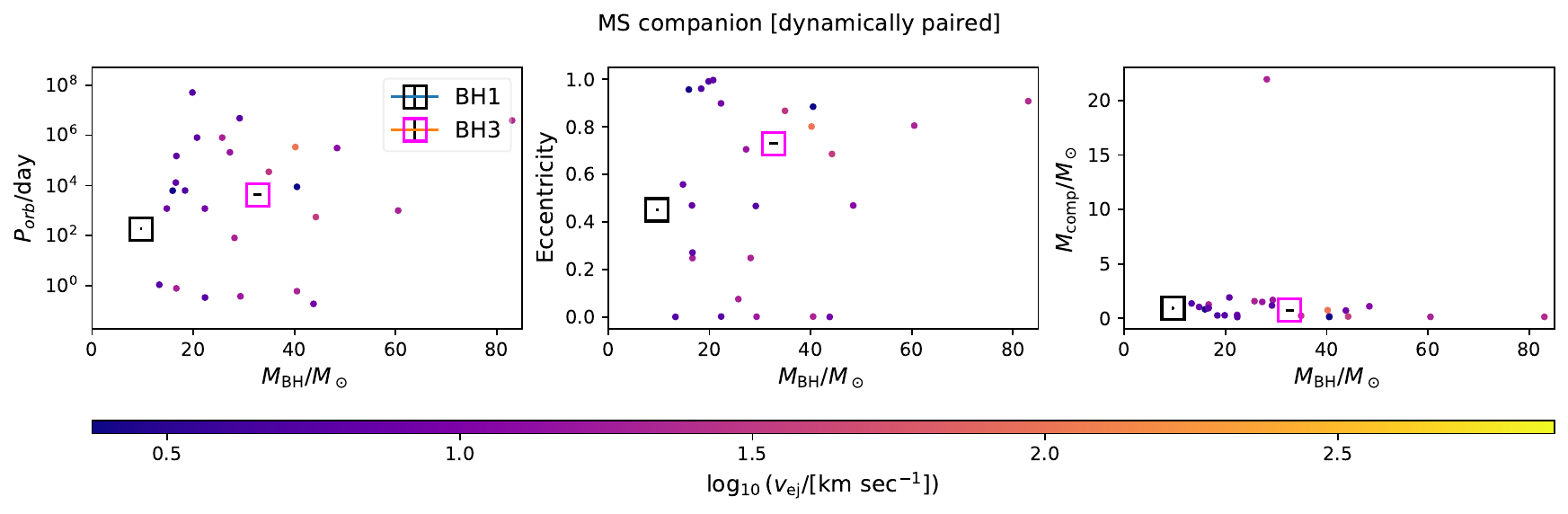}\\
\vspace{-1.3cm}
\includegraphics[width = 0.99\linewidth, angle=0.0]{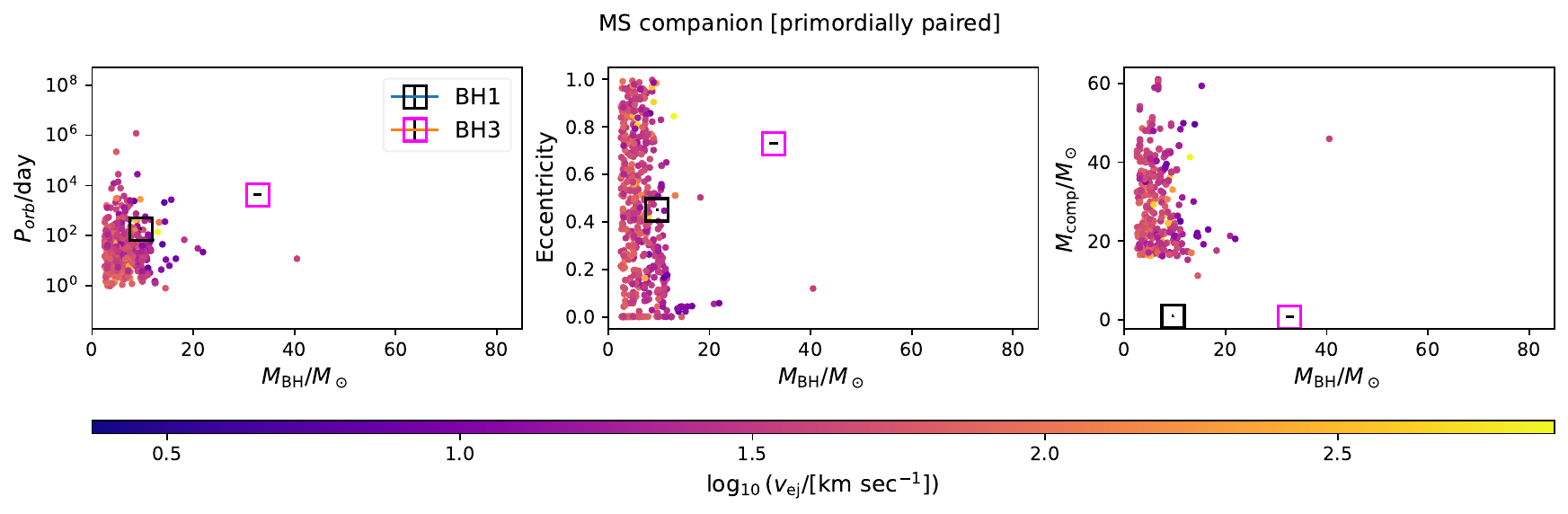}\\
\includegraphics[width = 0.99\linewidth, angle=0.0]{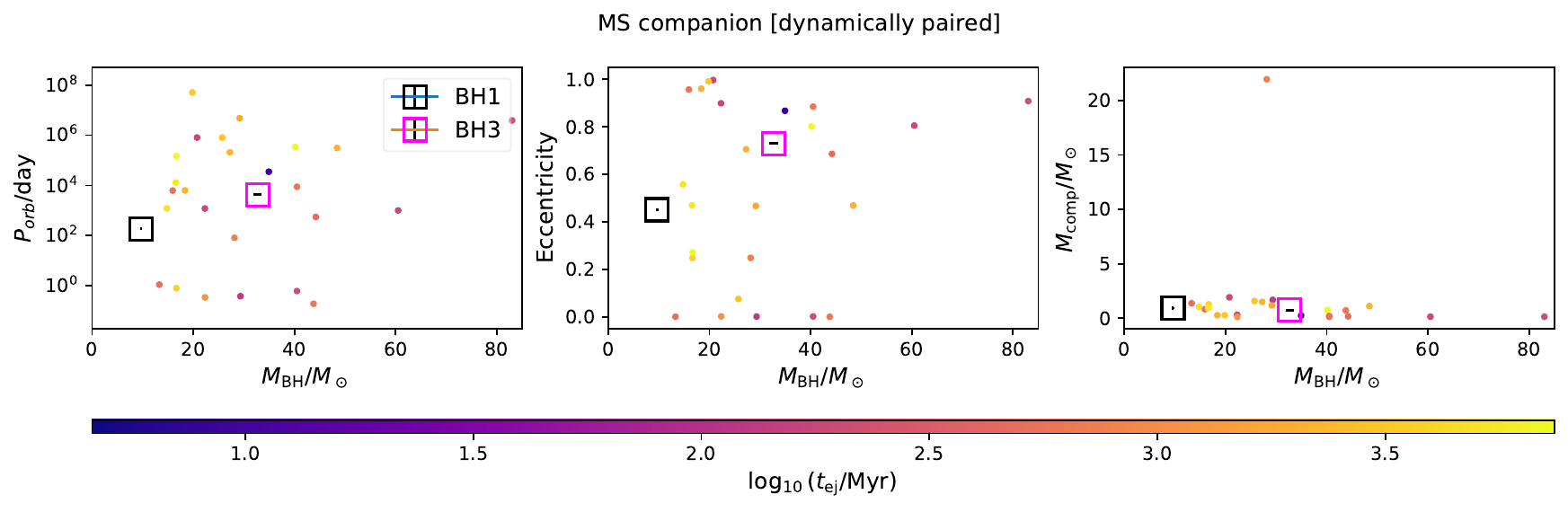}\\
\vspace{-1.3cm}
\includegraphics[width = 0.99\linewidth, angle=0.0]{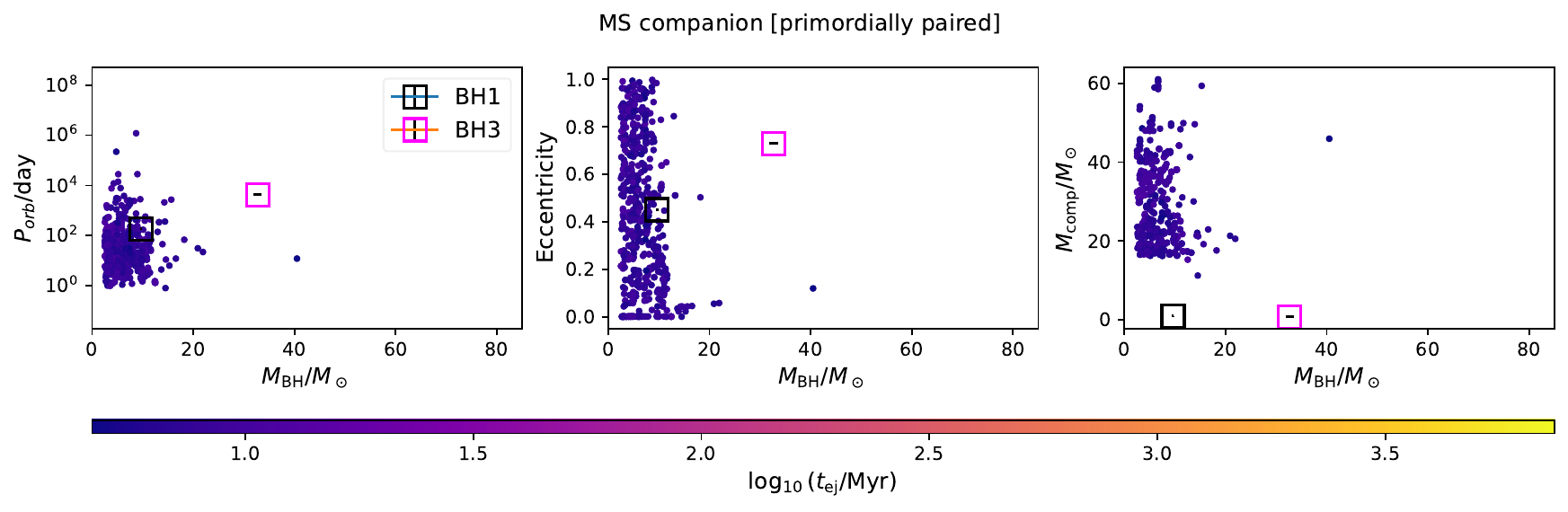}
\caption{Demographics of the BH-MS star binaries that have escaped into the galactic field
from the present evolutionary model star cluster grid.
In each row, the filled circles in the panels show the mass of the
BH member ($\mbh$; X-axes) against the orbital period ($\porb$; left panel),
eccentricity (middle), and companion-star mass ($\mcomp$; right)
of these model BH-MS binaries. The binaries are plotted separately, based
on the assembly sub-channel (primordial or dynamical), as indicated in each row's title.
The data points are colour-coded according to the velocity, $\vej$ (first and second row),
and time, $\tej$ (third and fourth row), of the binary's ejection.
All plotted values correspond to the event of the BH-MS binary crossing the instantaneous tidal
radius of its parent cluster. For comparison, the observed BH-MS binary candidates
Gaia BH1 and Gaia BH3 are shown in each panel (empty squares).
For these observed data points, the colour coding is not followed
and their error bars are generally invisible due to the scale of the figure axes. (The size
of the squares are chosen for legibility and does not represent
measurement uncertainties.)
	}
\label{fig:bhms}
\end{figure*}

\begin{figure*}
\centering
\includegraphics[width = 0.95\linewidth, angle=0.0]{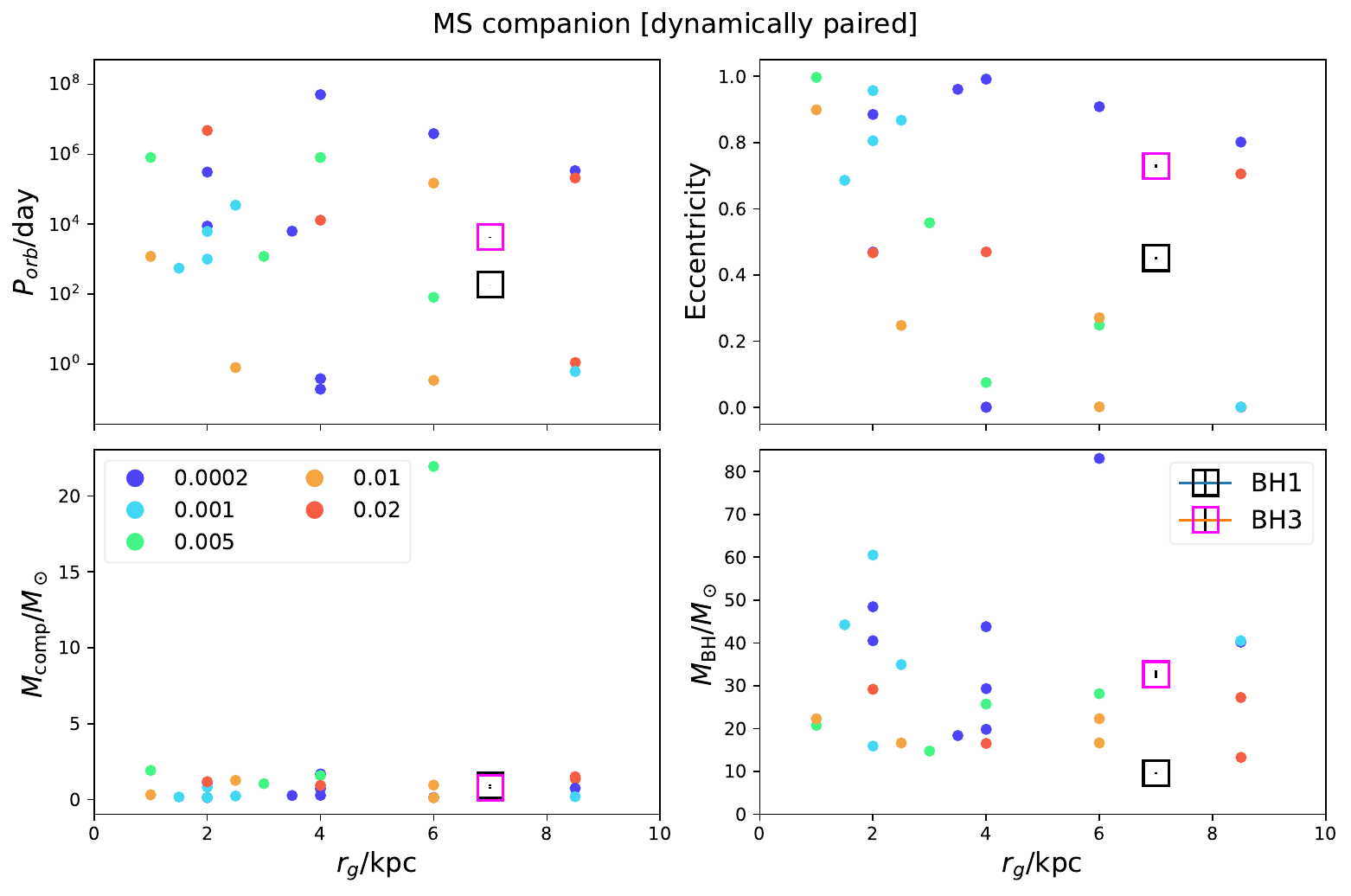}\\
\includegraphics[width = 0.95\linewidth, angle=0.0]{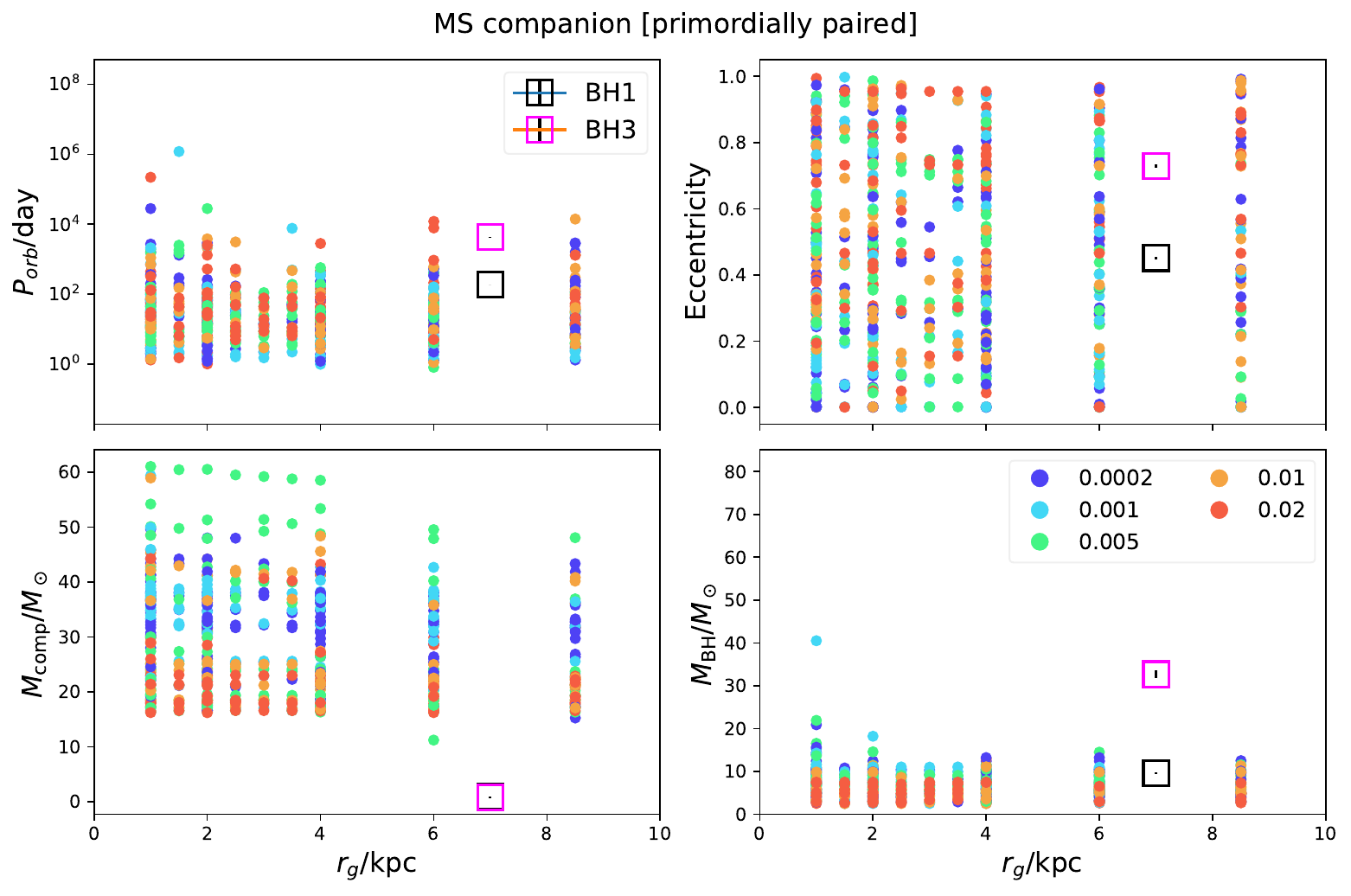}
\caption{Demographics of the BH-MS star binaries that have escaped into the galactic field
from all of the present evolutionary model star clusters. The descriptions of the quantities
plotted are the same as in Fig.~\ref{fig:bhms}, except that the binaries are plotted
against the galactocentric distance, $r_g$ (X-axes), of their respective parent cluster.
The data points are colour-coded according to their respective parent cluster's metallicity (legend).}
\label{fig:bhms_rg}
\end{figure*}

\subsection{Formation of black hole-main sequence binaries}\label{bhms}

The current models also produce detached (\ie, non-mass-transferring)
BH-star binaries that get ejected from the cluster and escape to the
galactic field. The formation mechanisms and the overall properties of these field BH-star binaries are similar
to those found in earlier star cluster computations by the author. The discussion is, therefore, not
repeated here and the reader is encouraged to consult Ref.~\cite{Kotko_2024} for all the details.
Fig.~\ref{fig:bhms} shows the ejected black hole-main sequence (hereafter BH-MS) binaries from only 
the present model grid, distinguishing between dynamically paired and primordially paired systems.
Consistently with the previous work (see also Refs.~\cite{DiCarlo_2023,MarinPina_2024b}), the parameters of the dynamically
paired field BH-MS binaries from the clusters encompass those of the observed, detached BH-MS binary candidates
Gaia-BH1 \citep{Chakrabarti_2023,ElBadry_2023} and Gaia-BH3 \citep{GaiaBH3_2024} in the MW field.
However, the parameters of the primordially paired ones disagree rather strongly with these observed
binaries' parameters. For Gaia-BH1, the disagreement is mainly due to its low component-star mass
that the primordially paired BH-MS binaries hardly possess (see Fig.~\ref{fig:bhms}). For Gaia-BH3,
the disagreement with the primordially paired population is due to the binary's low component-star mass, high BH mass,
and long orbital period.

The total formation efficiency of field BH-MS binaries containing a low mass ($<3\Ms$) MS companion from the
present models is $8.4\times10^{-6}\Ms^{-1}$. Restricting to only solar-mass-like ($0.7\Ms-1.1\Ms$)  
companions, the formation efficiency is $2.1\times10^{-6}\Ms^{-1}$. All of the escaped BH-MS binaries containing a
low mass MS companion are dynamically paired (c.f. Fig.~\ref{fig:bhms}). As discussed in
Ref.~\cite{Kotko_2024}, the rather low formation efficiency of escaped BH-star binaries
is due to the fact that the dynamically interacting BH-star systems inside the clusters favourably
undergo exchange interactions to become a BBH.

Fig.~\ref{fig:bhms_rg} shows the escaped BH-MS binaries plotted against the parent clusters' $r_g$.
As seen, dynamically formed Gaia-BH-type binaries can be ejected at all $r_g$s, \ie, from both
tidally under-filled and strongly stripped clusters. Notably, the high BH mass of Gaia-BH3
restricts its formation environment to mainly low metallicity clusters, as consistent
with the observed environment of this binary \citep{GaiaBH3_2024}.

\section{Summary and discussions}\label{summary}

This study investigates the role of tidal stripping of star
clusters on their production of BBH mergers. The concern regarding tidal stripping arises
since YMC/OC systems, while conceived as a significant resource of dynamically assembled
BBH mergers \citep{Banerjee_2010,Banerjee_2017,DiCarlo_2020}, are vulnerable to dissolution
by their host galaxy's tidal field (Sec.~\ref{intro}). To that end, a dedicated grid of
evolutionary model star clusters of initial mass $\mcl(0)=3.0\times10^4\Ms$, initial
sizes, $\rh(0)=1$, 2, and 3 pc, and metallicities $Z=0.0002$, 0.001, 0.005, 0.01, and 0.02
is computed. The grid comprises a total of 95 models that are subjected to a MW-like
external galactic potential at galactocentric distances ranging over
$1{\rm~kpc}\leq r_g \leq 8.5{\rm~kpc}$ (Sec.~\ref{grid}; Fig.~\ref{fig:grid}).
The model clusters are evolved with the direct N-body code $\nbseven$ that incorporates
stellar-remnant NS and BH formation and PN treatment of in-cluster NS- and BH-containing
binaries (Sec.~\ref{nbsims}). All clusters are evolved until late dynamical ages; at least until
the depletion of the BHs inside them and often until the cluster's dissolution.

Even the clusters at the smallest $r_g$, \ie, those experiencing the strongest galactic tidal
field can produce dynamical BBH mergers, despite such clusters' short lifetime (Sec.~\ref{gwmrg}).
This is because despite its destructive effect, tidal stripping aids the
dynamical relaxation of a cluster, which process is the key to the formation of dynamical BBH mergers
from them (Sec.~\ref{intro}). In particular, the central segregation and core formation of the cluster-retained BHs 
take place even at small $r_g$s where the cluster is rapidly dissolved by the external field
(Fig.~\ref{fig:rhratio}). Also, despite the expansion of the cluster due to BH heating,
the clusters tend to remain gravitationally bound and compact at all $r_g$ (Sec.~\ref{evolve};
Figs.~\ref{fig:conc},\ref{fig:fill}).
In fact, the merger efficiency, $\etamrg$, tends to increase with strengthening external potential,
once the tidal stripping starts to become important (Fig.~\ref{fig:etamrg}). 
The overall forms of the present-day-observable primary mass and mass ratio distributions of the BBH mergers
and the form of the merger rate evolution with redshift are similar for the tidally disrupted and the
tidally under-filled clusters (Fig.~\ref{fig:m1dist}). Despite significant tidal stripping,
the clusters are capable of producing Gaia-BH-like detached BH-MS binaries dynamically,
and eject them in the galactic field (Figs.~\ref{fig:bhms},\ref{fig:bhms_rg}).

A limitation of the present study is the use of a static, axisymmetric, MW-like external galactic
potential and a fixed initial cluster mass, which compromise the generality of the study.
These limitations are partly due to the tediousness of direct N-body computations, an approach
that is indispensable for properly treating the transition of a cluster into a tidally dissolving
phase (Sec.~\ref{nbsims}). However, these specifics of the model grid have also provided the opportunity  
to systematically explore the impact of external potential on GW mergers from clusters,
which is the main question that this study intends to address. 
As discussed in Sec.~\ref{nbsims}, the choice of $\mcl(0)=3\times10^4\Ms$ is driven by the fact that
GR mergers from initially pc-scale clusters (the initial scale length being motivated by observations
and theoretical models of massive star cluster formation) begin to become significant from around this mass.
While a more massive cluster would produce a larger number of mergers, it would generally be less affected
by the variation of $r_g$. That way, the present choice of cluster mass serves as an optimal probe
of the interplay between a cluster's internal relaxation  
and tidal stripping and its implications on GR merger production.

Another potential drawback is the representation of the model
grid with only one evolutionary model cluster per grid point. At present, this limitation stems
primarily from the high computational cost of direct N-body simulation of models initiating with $\mcl(0)\approx3\times10^4\Ms$
(membership $N(0)\approx5\times10^4$) and containing primordial binaries (see Sec.~\ref{nbsims}).
As each model cluster typically produces a few GR mergers, the mergers' delay times and
rate distributions are prone to
statistical uncertainties that are not represented in the current results. Qualitative agreements in trends
are nevertheless apparent between the different tidal-filling cases and metallicities considered
(Figs.~\ref{fig:m1dist} and \ref{fig:etamrg}).

Notably, a model cluster grid that includes variation in $r_g$ in a 
MW-like external potential has also been presented in Ref.~\cite{Kremer_2020}. This work comprises
Monte Carlo N-body evolutionary simulations \citep{Henon_1975} of a grid of GC-like model clusters (the `CMC Cluster Catalogue')
that are $\sim10$ times more massive than the present models.
Expectedly, although these models do exhibit a stronger
disruption with decreasing $r_g$ (see their Table A1), they are generally more resilient to tidal disruption.
Accordingly, these models do not show any particular trend in the BBH-mergers (and compact binary formation)
from them (see their Table A5). It is, however, noteworthy that CMC model computations
(Monte Carlo N-body simulations in general) need to be
halted well before the two-body relaxation timescale of a cluster becomes comparable to the cluster's dynamical
timescale (see Ref.~\cite{Chatterjee_2017a} and references therein) as the cluster continues to deplete.
That way, the signatures of enhanced tidal stripping may not be fully captured in the CMC models.

The BBH-merger formation efficiencies of the CMC cluster models are typically
$\sim10^{-4}\Ms^{-1}$, \ie, $\sim10$ times higher than those of the current models. This difference is likely
due to the much higher, GC-like masses of the CMC models and also due to their assumption of the
`rapid' remnant mass model, as opposed to the delayed model in the present computations. The
rapid model causes a higher BH retention in clusters than its delayed counterpart \citep{Banerjee_2020}. Interestingly,
more recent direct N-body computations of star clusters of masses comparable to the CMC models
(the `DRAGON-II' simulations \cite{ArcaSedda_2024c}) yield merger efficiencies $\sim10^{-5}\Ms^{-1}$, \ie,
similar to those of the present models.

Recently, several authors have considered star cluster models with extremely compact, $\approx0.1-0.3$ pc,
initial sizes (\eg, Refs.~\cite{Kumamoto_2020,DiCarlo_2020,Rastello_2021}). By evolving from such a compact initial size,
a cluster would potentially be much less affected by the external field. However, such initial scales
represent those of gas-embedded clusters \citep{Krumholz_2019}. It can be expected that the gas eventually gets expelled 
due to radiation pressure and mechanical (\eg, wind) feedback from the cluster's massive stars \citep{Krumholz_2009}.
With a typical $\approx30$\% star-to-gas fraction \citep{Brinkmann_2017},
or with even smaller star formation efficiency \citep{Parmentier_2013,Shukirgaliyev_2017},
the gas expulsion would expand the cluster to a pc-scale size
within a few Myr age, if not dissolve the cluster completely \citep{Calura_2015,Brinkmann_2017,Banerjee_2018b}.
However, there is ambiguity regarding
whether gas can always be successfully expelled from a newborn embedded cluster, or
whether it is possible to convert all of the gas into stars \citep{Longmore_2014,Krumholz_2014,Banerjee_2018b}.
Observationally, YMCs and OCs are typically gas free and are of parsecs length-scale \citep{PortegiesZwart_2010},
consistent with the initial conditions adopted here.

Furthermore, star clusters are observed to be born inside dense molecular clouds,
where the star formation is heavily sub-structured and/or distributed along gas filaments \citep{Andre_2014,Krumholz_2019}.
It is widely conjectured that the present-day-observed, near monolithic YMCs are hierarchically assembled
from such an irregular, gas-embedded configuration by undergoing a violent relaxation phase.
A YMC is formed if the stellar system survives the violent relaxation, gas dispersal, and potential stripping
by external tidal field \citep{Longmore_2014,Banerjee_2018b}.
The monolithic, virial, gas-free initial cluster configuration, as adopted here, therefore implicitly
preselects those systems that survive the cluster assembly phase. This corresponds to an effective
cluster formation efficiency (hereafter CFE) for YMCs, whose estimate is rather ambiguous
in the literature \citep{Pfeffer_2019,Reina-Campos_2022}. Merger rate estimates are dependent on the CFE, but since relative
rate distributions are mainly considered in this work (see Fig.~\ref{fig:m1dist} and the
associated description), the CFE value is irrelevant for the present results.
Also, $\etamrg$ is, by definition, independent of CFE. The present results, however,
implicitly assume that the CFE remains constant with redshift and $r_g$, as
commonly adopted in the literature due to the lack of estimates of such dependencies.

The early hierarchical assembly phase would cause an enhanced
mass segregation at the young age of the cluster \citep{Banerjee_2015,Dominguez_2017}. However, since the mass segregation
timescale of the BH-progenitor stars in such clusters is typically $\lesssim1$ Myr \citep{Spitzer_1987,Heggie_2003},
an initially mass-segregated state of the models (as opposed to the presently assumed
initially unsegregated state) would unlikely cause qualitative
differences in the clusters' evolution and their BBH-merger outcomes. The assembly phase
can also potentially cause a runaway growth of an intermediate-mass black hole,
but this process is probable in a significantly more massive stellar system \citep{Rantala_2024,Lahen_2025}. 
In summary, it can be expected that the complex star formation morphology and assembly phase
of YMCs are unlikely to qualitatively alter the current inferences.

It is nevertheless necessary to study the effect of galactic tidal field with additional cluster masses
(both lower and higher masses). This will help to convolve the effect with a cluster mass function.
It would also be worthwhile to consider cluster evolution under the influence of
the potential of an irregular and asymmetrical field mass distribution, such as those obtained
from high-resolution structure formation simulations (\eg, Refs.~\cite{Bruel_2024,Bruel_2025}).
Such studies will be taken up in the near future.

This work demonstrates the
importance of incorporating external galactic fields in cluster evolution calculations,
especially in connection to their GW-merger formation. It shows that strong tidal stripping
may not quench or significantly suppress the formation of dynamical GW mergers in moderately
massive YMCs and OCs.

%\begin{figure*}
%\centering
%\includegraphics[width = 10.0 cm, angle=0.0]{}
%\includegraphics[width = 10.0 cm, angle=0.0]{}
%\caption{}
%\label{fig:}
%\end{figure*}

%
%\nocite{*}

\section*{Data availability}

The simulation data are available upon reasonable request to the author.

\begin{acknowledgments}
The author (SB) thanks the reviewer for their constructive criticisms and suggestions that have
helped to improve the paper.
SB acknowledges funding for this work by the Deutsche Forschungsgemeinschaft
(DFG, German Research Foundation) through the project ``The dynamics of stellar-mass black holes in
dense stellar systems and their role in gravitational wave generation''
(project number 405620641; PI: S. Banerjee).
All $N$-body simulations have been carried out on the {\tt gpudyn}-series GPU computing servers containing
NVIDIA Ampere A40 and NVIDIA RTX 2080 GPUs, located at the AIfA, University of Bonn.
The {\tt gpudyn} servers have been sponsored by the above-mentioned DFG project and the HISKP. 
SB is thankful to the IT teams of the AIfA, HISKP, and University of Bonn HPC center for their generous support.
\end{acknowledgments}

\bibliography{bibliography/biblio.bib}

\appendix

\section{Additional illustrations}\label{figmore}

\begin{figure*}
\centering
\includegraphics[width = 0.99\linewidth, angle=0.0]{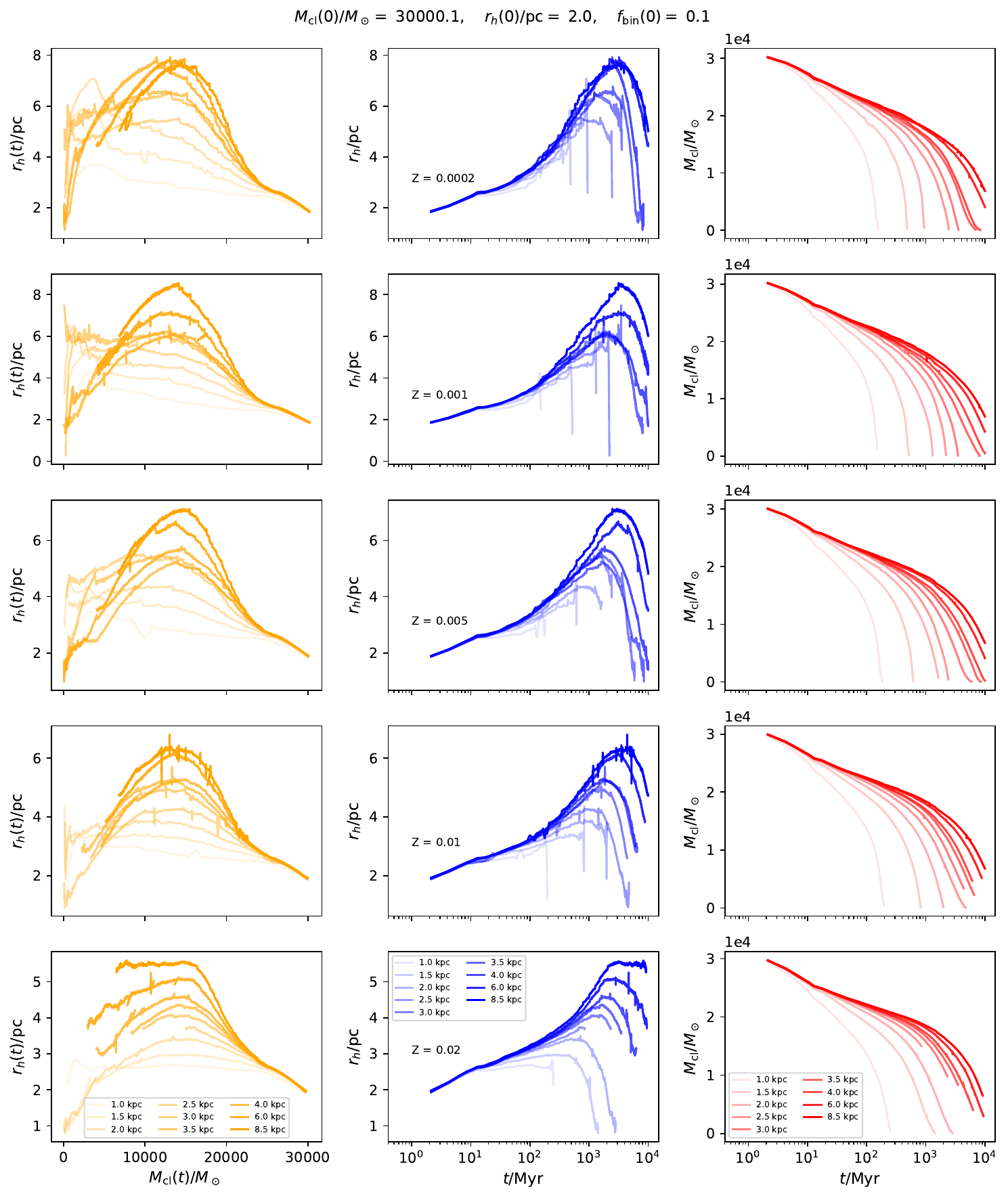}
	\caption{Time evolution of cluster size (half mass radius, $\rh$; middle column), total bound mass,
	$\mcl$ (right column), and the mass-size relation (left column) for the $\rh(0)=2$ pc model clusters.
        The models with varying metallicities are plotted along distinct rows, while those at different $r_g$s
	are distinguished by varying shades as indicated.}
\label{fig:mcl-rh-t}
\end{figure*}

\begin{figure*}
\centering
\includegraphics[width = 0.99\linewidth, angle=0.0]{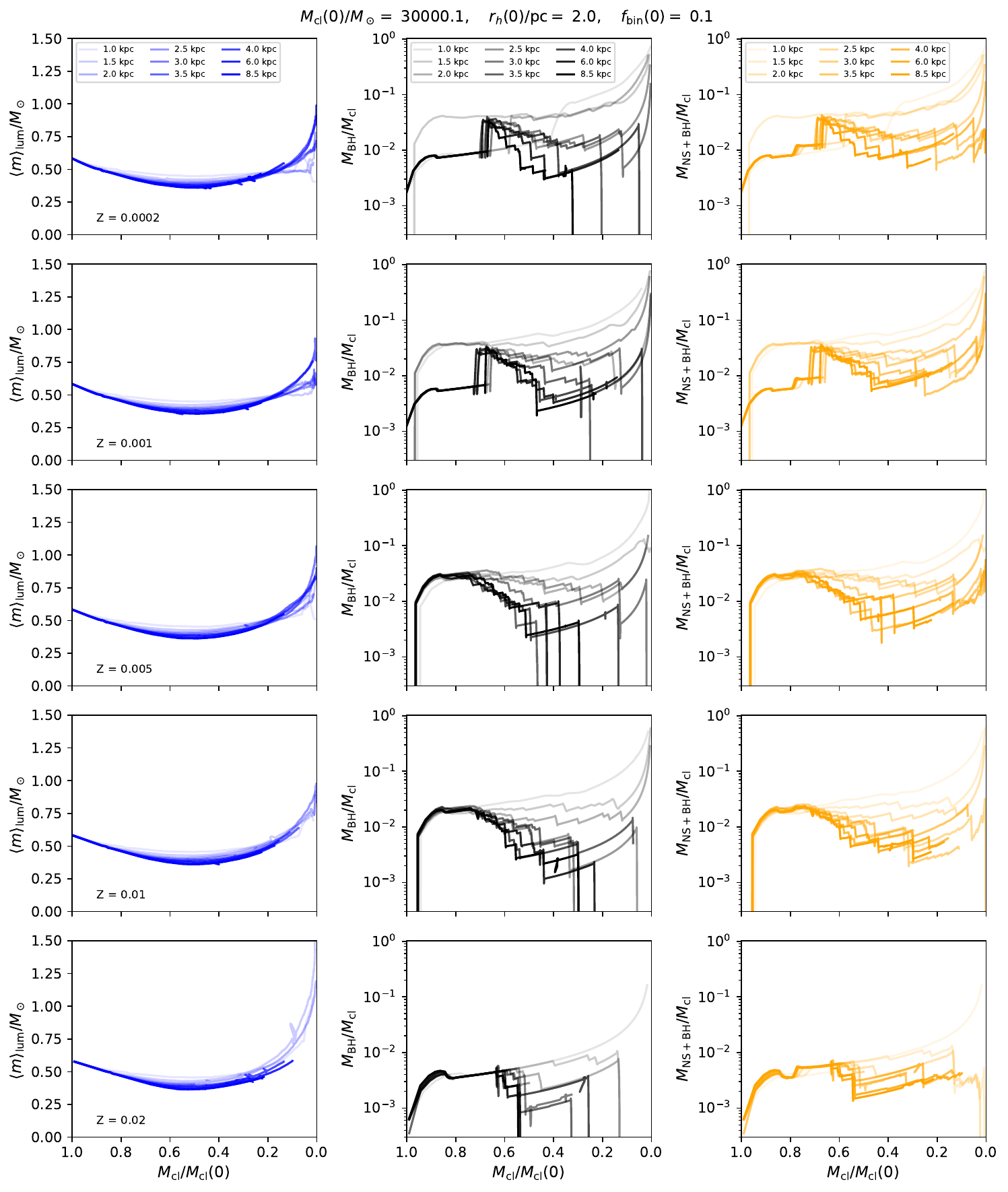}
	\caption{Evolution of the mean mass of the luminous members (left column), the mass fraction of black holes
	(middle column), and the mass fraction of dark remnants (\ie, of black holes and neutron stars combined;
	right column) with instantaneous total bound mass (X-axis), for the $\rh(0)=2$ pc model clusters.
        The models with varying metallicities are plotted along distinct rows, while those at different $r_g$s
	are distinguished by varying shades as indicated.}
\label{fig:darkfrac}
\end{figure*}

\end{document}